\documentclass[lettersize,journal]{IEEEtran}
\usepackage{amsmath,amsfonts}
\usepackage{algorithmic}
\usepackage{algorithm}
\usepackage{array}
\usepackage{subfig}
\usepackage{textcomp}
\usepackage{stfloats}
\usepackage{url}
\usepackage{verbatim}
\usepackage{graphicx}
\usepackage{cite}
\usepackage{xcolor}
\usepackage{color,soul}

\begin{document}

\title{Scaling up Energy-Aware Multi-Agent Reinforcement Learning for Mission-Oriented Drone Networks with Individual Reward}

\author{Changling Li, Ying Li,~\IEEEmembership{Member,~IEEE}, 
\thanks{Manuscript received April 2 2024; revised August 15 2024 and 8 November 2024; accepted 2 December 2024.}
\thanks{Changling Li is with the Department of Computer Science, ETH Zurich, 8092, Zurich, Switzerland (email: lichan@student.ethz.ch).}%
\thanks{Ying Li is with the Department of Computer Science, Colby College, 04901, Waterville, ME, USA (email: yingli@colby.edu).}
}

\markboth{IEEE INTERNET OF THINGS JOURNAL.}%
{Shell \MakeLowercase{\textit{et al.}}: A Sample Article Using IEEEtran.cls for IEEE Journals}

\maketitle

\begin{abstract}
Multi-agent reinforcement learning (MARL) has shown wide applicability in collaborative systems such as autonomous driving and smart cities for its ability of learning through interaction. With the recent development of drone networks, researchers have also applied MARL to address the trajectory planning problems. However, the dynamic environment and the limited battery capacity are still challenging for using MARL to achieve efficient collaborative task execution. In this paper, we propose an energy-aware MARL model as an attempt to tackle these challenges, leveraging Deep Q-Networks (DQN) with \emph{individual reward functions} driven by the task execution progress and the remaining battery of drones. We conduct a set of simulation studies for the proposed mode and compare it with the shared reward MARL~\cite{Li2022MARL} to explore the impact of credit assignment in MARL. The results indicate that our proposed model can achieve at least 80\% success rate regardless of the task locations and lengths. Similar to the shared reward mode, the individual reward mode can achieve a better success rate when the task density is high, and it can hit nearly a 100\% success rate when task density gets close to 40\%. The true advantage of our proposed model with individual reward is revealed when scaling up the environment. The comparison to the shared reward MARL shows that the our proposed model is more robust towards the change of the environment size and agent numbers. It can achieve higher success rate with fewer steps due to the clarity of the goal which improves energy efficiency even better.
\end{abstract}


\section{Introduction}
\IEEEPARstart{M}{ulti-agent} reinforcement learning (MARL) is an extension of single-agent reinforcement learning (RL) where multiple agents follow the Markov decision process to form control policies for their purpose~\cite{mnih2015human}. In a MARL, agents learn the control policies through the experience of interacting with the environment and other agents~\cite{hong2017deep}\cite{oroojlooyjadid2019review}. Depending on the application scenarios, various MARL models are created to achieve cooperation~\cite{yao2019collaborative}, competition~\cite{tampuu2017multiagent}, or a mix of both~\cite{bucsoniu2010multi}\cite{lowe2017multi}\cite{chen2020delay}. With its its ability to solve sequential decision-making problems and the recent development in RL~\cite{schaul2015prioritized}~\cite{haarnoja2018soft}~\cite{fujimoto2018addressing}~\cite{li2024roer}, MARL has shown wide application in diverse fields including smart cities~\cite{vazquez2019multi}, multi-robot systems~\cite{duan2012multi}, traffic control~\cite{zhang2019integrating}, and planning \cite{9537638}, \cite{10153615}.

Drone network is another field where MARL shows successful application. Drone networks have become increasingly popular recently because of their wide applications in delivery~\cite{pugliese2020using}, surveillance\cite{10606397}, mobile edge computing\cite{10606273}, sensor network\cite{10445189}, and metaverse\cite{10415630}. Researchers have explored the application of MARL in a swarm of drones for providing the next generation of cellular networks~\cite{hu2020cooperative}. However, our focus, mission-oriented drone networks, has received little attention for adopting MARL. In mission-oriented drone networks, a mission can consist of several tasks. The task locations are often random and the lengths may vary for different tasks, which creates a dynamic environment. An example is the inspection of wind turbines where drones are used to inspect the surface damage~\cite{shihavuddin2019wind}. Depending on the damage level, the operation time required can be different and multiple drones may be needed to inspect one turbine. We note that this type of problem differs from Vehicle Routing Problem (VRP). VRP commonly involves tasks with fixed positions and the goal is to identify a route for passing through points~\cite{toth2002vehicle}. In contrast, our defined problem concerns tasks with dynamic locations and various lengths. The tasks have non-binary lengths and may require multiple drones to collaboratively execute. This makes the traditional VRP algorithms inefficient and infeasible. In addition, the current battery technology only allows drones to operate for up to 40 minutes~\cite{hashemi2019new}. These challenges require more robust approaches to generate efficient trajectory planning and task execution strategies. The ability of MARL to learn by interacting with the environment and form control policies that work for dynamic environment makes it a suitable solution to tackle the proposed problems.

\IEEEpubidadjcol

However, the applicability of MARL also comes with the challenge of credit assignment~\cite{du2021survey}. Agents in a cooperative MARL receive the same rewards at each episode determined by the shared reward function. This reward system aims to foster collaboration among agents with the cost of blurring each agent's contribution to the goal achievement, resulting in inefficiency.  

To explore the application of MARL in mission-oriented drone networks and address the reward assignments problem, we propose a collaborative execution algorithm based on MARL with an individual reward mode for a mission-oriented drone network. %
Different from the shared reward mode, in which we form a fully cooperative stochastic game and the reward received by each agent at each step is the same, in the individual reward mode, we form a mixed game, and each agent receives a reward based on its own action at each step. The application scenarios consist of multiple independent tasks with various lengths at different locations. Each task requires more than one time step to finish, and a mission is completed when all tasks are finished. A drone can execute more than one task, and multiple drones can co-execute a task. Considering the constraint of the battery capacity, our reward functions are formulated based on both task execution process and remaining battery power. To our best knowledge, our exploration is one of the pioneer studies focusing on the application of MARL to cooperative mission execution of drone networks with the consideration of battery capacity, the impact of reward assignment and the applicability to embedded systems. Our proposed energy-aware MARL model with individual reward is robust towards both changes in task lengths and locations and shows consistent good performance across different numbers of tasks and sizes of environments compared to the shared reward scenario.

The rest of the paper is structured as follow. Section~\ref{related work} presents the related work in MARL and reward study and lists the contribution of this work. Section~\ref{system model}
describes the detailed energy model of drones in our study. Section~\ref{proposed model} exhibits our proposed model and reward formulations. The proposed model is then evaluated on scalability and robustness in Section~\ref{performance eval}. We discuss the advantages and limitations of our work in Section~\ref{conclusion} and outline our future work. 

\section{Related Work}
\label{related work}
Past works have explored the applicability MARL models to solve the trajectory planning problem in dynamic environments and achieve higher utility of drone networks. Wang~\textit{et al.} focus on trajectory planning of multiple drones without collisions and they apply actor-critic RL with a convolutional neural network to process observations~\cite{wang20213m}. Cao~\textit{et al.} propose a hierarchical pipeline combining a genetic algorithm for global multi-robot planning and reinforcement learning for local navigation to achieve efficient multi-agent navigation \cite{10309976}. Chen~\textit{et al.} concern the quality of service provided by the drone networks and employ a MARL-based framework to determine the optimal trajectory of each drone in a distributed manner~\cite{chen2021multi}. Similarly, Zhao~\textit{et al.} apply Multi-agent Deep Deterministic Policy Gradient (MADDPG) algorithm to optimize the utility of drone network while guaranteeing the quality of service by modeling the joint trajectory design and power allocation issue as a stochastic game~\cite{zhao2020multi}. MADDPG is also utilized by Wang~\textit{et al.} to maximize the geographical fairness among all the user equipment and the fairness of each drone's user equipment load and minimize the overall energy consumption of user equipment~\cite{wang2020multi}. Khalil~\textit{et al.} consider the path planning problem of drone networks as an economic game and created a MARL algorithm inspired by economic transactions to distribute tasks between drones~\cite{khalil2021efficient}. We note that all the aforementioned work limits their studies of applying existing RL algorithms to only the trajectory planning tasks while our work focuses on creating a MARL framework for comprehensive workflow of task allocation, task execution planning and path planning. The exploration of MARL in task-oriented drone networks is limited. He~\textit{et al.} applied deep RL in multi-agent systems for task-oriented communication and forms a modular design for data transmission and execution~\cite{10183796}. Liu~\textit{et al.} uses MARL for drone networks to solve the dynamic task allocation problems in heterogeneous UAVs~\cite{9980391}. These two works, similarly, also limit their application of MARL to only a subset of the complete workflow. In addition, they lack details in formulating the learning goals, which are decided by the reward function. Our work bridges the gap and studies the reward formulation. We primarily focus on improving energy efficiency through the formulation of the reward function, which has not been explored before in applications. In addition, different from other works applying MARL to drone networks, we consider the tasks in our simulation to have a non-binary length.

Reward formulation controls the performance of the MARL system. However, it also faces the issue of credit assignment, which refers to the incorrect reward assignment associated with the contribution of the agents. Various methods have been explored to solve this problem. Wang~\textit{et al.} propose a local reward approach that reflects individual agents' contribution to the global goal to overcome the ambiguity issue caused by the inaccurate reward assignment~\cite{wang2020shapley}. They form a Shapley Q-value deep deterministic policy gradient algorithm that learns decentralized policies with centralized critics. Other works explored the separation between individual goals and global goals. Yang~\textit{et al.} consider the cooperative multi-agent control problem consists of two stages~\cite{yang2018cm3}. Agents are trained to achieve individual goals prior to learning of the cooperation. They derived a policy gradient with a reward function for individual reward assignment and used function augmentation to bridge value and policy functions across the stage. Sheikh~\textit{et al.} focus on maximizing individual and global rewards through the proposed method Decomposed MADDPG, which uses a global critic and a local critic~\cite{sheikh2020multi}. Du~\textit{et al.} combine the centralized critic with the individual reward shaping to generate diversified behaviors for each agent in a cooperative MARL~\cite{du2019liir}. Here, we differ our study from the aforementioned work by considering practical usage and scenario-specific reward formulation. This work investigate the impact of credit assignment within mission-oriented drone networks. We detailedly analyze the performance improvement of our method compared to the shared reward from the perspective of credit assignment. We propose a cooperative task execution model for drone networks driven by the battery capacity of drones based on deep Q-networks (DQN). Our method enables effective task assignment, trajectory planning, and task execution strategy for drone networks by formulating appropriate reward functions. In summary, the main contributions of this work are:
\begin{itemize}
\item We consider a practical scenario where the drones need to stay on a spot for multiple time steps to execute the task. The tasks in our simulation have non-binary lengths and need more than one time step to finish. We use the wind turbine inspection as an illustrative example and provide the simulation environment implementation. However, we note that the task locations and task lengths of a mission in our study are non-predetermined which ensures the applicability beyond our example environment.
\item The formulation of the reward function for DQN is driven by the task execution progress and the battery level of all drones to enable the efficient task assignment, trajectory planning, and task execution. Our proposed method is evaluated in detail with the exploration of specific hyperparameters and task density analysis. To our best knowledge, it is one of the first few works that consider the battery capacity of drones in the creation of MARL models. 
\item We explore the impact of reward assignment issue in our simulation by forming both shared reward scenario and individual reward scenario. The performances of both cases are presented and analyzed in detail to show the advantage of individual reward mode in scalability and task completion efficiency. 
\end{itemize}

\section{System Model}
\label{system model}

Our research concerns the battery capacity of drones, and a detailed model of drone energy consumption is necessary to create a realistic simulation. We adopt the calculation method of energy consumption proposed by Stolaroff~\textit{et al.}~\cite{Stolaroff_nature_2018} in our experiment. During a mission, a drone consumes energy in the following three modes: forward mode, hovering mode, and execution mode. In forward mode, for each time step, the estimated minimum energy consumption rate, $P_{F}$ is defined by Equation~\ref{eq:forward},

\begin{equation}
P_{F} = \frac{(m_{drone} g + F_{drag})(v \sin\alpha + v_s)}{\eta},
\label{eq:forward}
\end{equation} 

\noindent where $m_{drone}$ is the total mass of a drone, including its battery and equipped facilities for task execution, $g$ is the gravitational constant, $F_{drag}$ is the drag force which considers the wind resistance, $v$ is the ground speed of the drone, $\eta$ is the overall battery efficiency of a drone, $\alpha$ is the pitch angle of a drone for steady flight, $v_s$ is the included velocity to achieve the trust to forward at the desired speed and can be calculated by the following Equation~\ref{eq:vs}, 

\begin{equation}
v_s = \frac{2 (m_{drone} g + F_{drag})}{\pi c D^2 \rho \sqrt{(v \cos\alpha)^2+(v \sin\alpha + v_s)^2}}.
\label{eq:vs}
\end{equation}   

\noindent $D$ here is the diameter of each drone rotor and $c$ is the number of rotors a drone has. 

As the drone switches to hovering mode, the energy consumption rate per time step, $P_{H}$, is defined by Equations~\ref{eq:phover},

\begin{equation}
P_{H} = \frac{(m_{drone} g + F_{drag})^\frac{3}{2}}{\eta\sqrt{\frac{1}{2}\pi cD^2\rho}},
\label{eq:phover}
\end{equation}

\noindent where $\rho$ is the density of air. 

During task execution, the energy consumption of a drone consists of power required for both hovering and supporting equipped facilities (e.g., camera, sensor, embedded neural network etc). For each time step, the energy consumed by the facilities carried by a drone is $P_{E}$. The value of $P_{E}$ depends on the type of facilities and their energy consumption rate.

Here, we specifically discuss the energy consumption of loading neural network in the embedded device to address that it does not counter our goal of achieving energy-efficient mission execution. In our proposed model, the system is trained in the base station and the drones only consume energy for predictions of the neural networks. Past work has shown that such consumption of energy can be rather small \cite{arnautovic2021evaluation}. Various techniques focusing on hardware development such as \cite{loni2019neuropower},\cite{lee2019neuro},\cite{kim2020energy} and \cite{verhelst2017embedded} have also been proposed to reduce the energy consumption of such which is, however, beyond the focus of this work. In addition, the neural network structure in our proposed system model is rather minimal as described in Section~\ref{proposed model} which controls the energy usage to only necessity.

\begin{figure}[tbp]
\centerline{\includegraphics[width=\linewidth,height=0.5\linewidth]{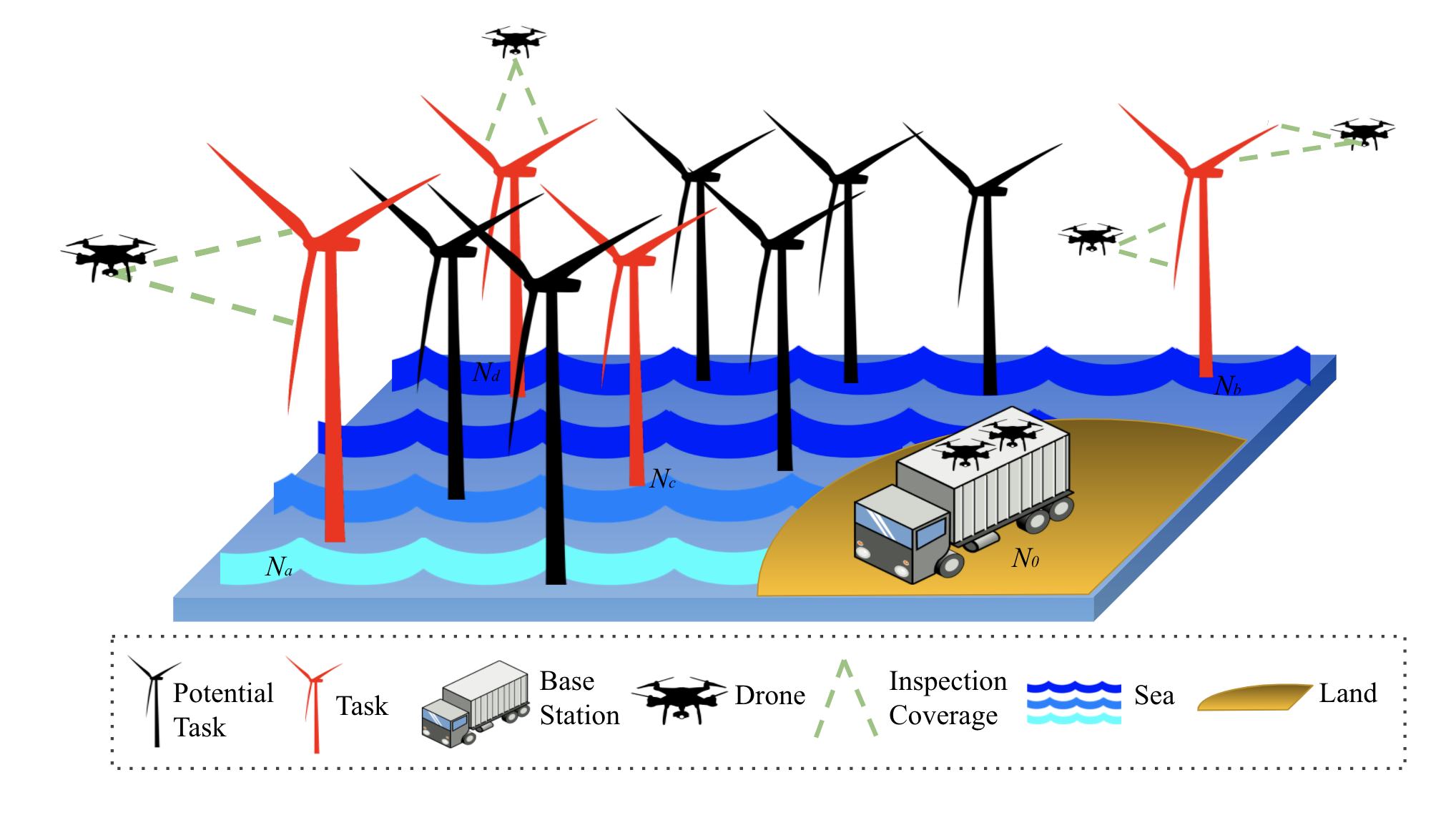}}
\caption{System Model for Task Execution in Mission-oriented Drone Networks}
\label{fig1:system model}
\end{figure}

In our simulation, We assume that all drones have the same configuration, carry the same facilities for task execution, and travel at a constant speed to simplify the calculation. Thus, $P_{F}$, $P_{H}$, and $P_{E}$ each has the same value for all drones.

With the energy model, we created our simulation environment as depicted in Fig.~\ref{fig1:system model}. The simulation consists of a set of trajectory points, $\mathcal{N}  = \{N_0, \dots, N_n\}$. The starting point of all drone trajectories is located at $N_0$, which is the base station. Other trajectory points $\{N_1,\dots, N_n\}$ are potential task locations. For each simulation, tasks are distributed as a subset of the trajectory points $\{N_1,\dots, N_n\}$. We consider that every trajectory point only has one task, and a task can only locate at one trajectory point. No task locates at the base station. Thus, for each simulation, there are $K$ tasks where $K\in [1, n]$. Each task at trajectory point $N_i$ needs $\mathcal T_i$ time step to finish where $\mathcal T_i \geq 1$. Fig.~\ref{fig1:system model} shows a scenario with $K=4$ number of tasks located at four out of ten trajectory points, $n=10$.

For simplicity but without losing generality, we set the drone number the same as the task number. We consider a mission-oriented network that a mission consists of $K$ tasks is assigned to a drone network with $K$ drones. We note that different number of drones and tasks is also possible using our proposed model as our proposed MARL model and environment model are separated and can be easily controlled by setting different numbers for drones and tasks. At the beginning of each mission, $K$ tasks appear at a subset of trajectory points, and $K$ drones are fully charged for task execution. A fully charged drone has a battery capacity of $B$. All drones start from the base location $N_0$ and fleet to trajectory points where tasks are located to execute tasks. A drone can decide the number of time steps spent on execution based on the current status of the environment and itself. It can execute a portion of a task, $t_i \in [0, \mathcal T_i]$, or multiple tasks. Multiple drones can also execute different portions of a task at the same time. We assume that portions do not physically overlap in this study. For example, each portion is part of a turbine to be examined. Depending on the reward formulation, drones either cooperatively or competitively complete the mission after launching. A mission is accomplished when all tasks are finished and all drones return to the base stations from their current locations.

The goal of the mission-oriented network is to execute all the tasks assigned to the network without sacrificing any drone. This proposes two constraints for successful mission completion: 1) all assigned tasks in one mission are finished, and 2) all drones have enough energy left to return to the base station. These two requirements are each summarized by formulae~\ref{eq:constraint1} and~\ref{eq:constraint2}.

\begin{equation}
\sum_{i=1}^{K} \tau_i = 0, 
\label{eq:constraint1}
\end{equation} 
\begin{equation}
B_k \geq B^k_{R},\forall k \in \mathcal{K},
\label{eq:constraint2}
\end{equation} 

\noindent where $\tau_i$ indicates the remaining length in time step of the task at trajectory point $N_i$. The valid range for $\tau_i$ is $[0, \mathcal{T}_i]$ where $\mathcal{T}_i$ is the task length at trajectory pint $N_i$ at the beginning of the mission. $K$ is the total number of trajectory points where tasks are located. As aforementioned that there is an equal number of drones as the number of tasks, $k \in \mathcal{K}$ where $\mathcal{K}=\{1,\dots, K\}$. $B_k$ indicates the remaining energy of the drone $k$. $B^k_R$ represents the energy required for the drone $k$ to return to the base station from the current location. It can be calculated by Equation~\ref{eq:b_return},

\begin{equation}
B^k_{R} = \frac{d(N_i, N_0) \times P_{F}}{v},
\label{eq:b_return}
\end{equation}   

\noindent where $d(N_i, N_0)$ indicates the Euclidean distance between the trajectory point $N_i$ where the drone $k$ is currently located and the base station $N_0$. In our simulation, the valid range for $B_k$ is $(-\infty,B]$. The negative value indicates how much energy a drone $k$ spent beyond the battery capacity $B$. With the presence of the negative $B_k$, even though all the tasks are finished, $\sum_{i=1}^{K} \tau_i = 0$, the mission is considered to be failed as the drone k does not return to the base station safely. In addition, we consider that drones do not execute tasks on the way back to the base station. We also assume that the drones in our simulation are intelligent as the new Skydio 2 model that can automatically avoid collision with other drones or objects. 

\section{Proposed Collaborative Execution Model Based on MARL and DQN}
\label{proposed model}

We propose a collaborative execution model based on DQN to address the issue of limited energy for task executions in mission-oriented drone networks using an individual reward function. The newly proposed model is an extension of the model proposed in~\cite{Li2022MARL}. In this section, we present the detailed formulation used by the newly proposed model. 

For our model, each drone has a DQN. Through interaction with other drones and the environment, they can find near-optimal trajectories and task execution strategies to complete each mission. The DQN in our model has a target network and a policy network. Both networks have two fully-connected layers with rectified linear unit activation functions. The output layer of the networks is a dense layer, and the loss function of the networks is a mean squared error. To increase the efficiency of training, we adopted the Adam optimizer. We also employed the experience replay model in our training process to break the dependency among observations and make the model more robust. 

One of the purposes of this study is to understand the impact of reward. We formulated the individual reward function and compared it with the model using shared reward function~\cite{Li2022MARL}. The shared reward function model lets all drones receive the same reward at each time step. Each drone's action and state affect the reward received by all the drones. In contrast, the individual reward function only allows each drone to receive a reward based on its own state and action. We expect the individual reward function to form a sense of competition during the task execution stage. However, the ultimate goal is still to achieve collaborative task execution, as each drone receives a collective reward when all tasks are finished. A detailed description of state-action space and control policy are discussed below.

\subsection{Action and State Space}
\begin{figure}[htbp]
\centerline{\includegraphics[scale=0.4]{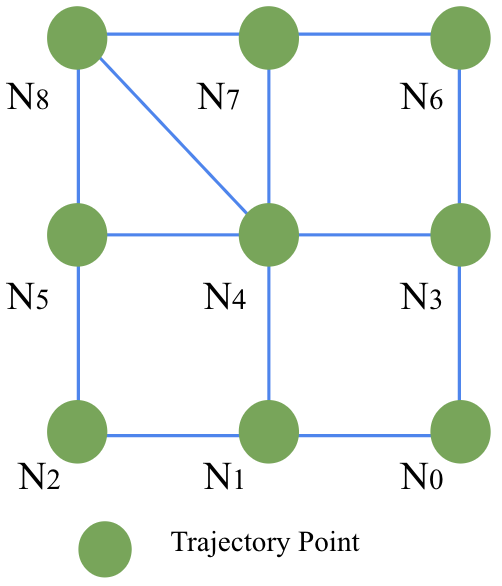}}
\caption{A grid pattern indication of endpoints for a drone located at trajectory point N4.}
\label{fig:gridPattern}
\end{figure}

We consider the environment of our simulation as a finite set of discrete trajectory points $\mathcal N$ similar to the simulation model in~\cite{hu2018reinforcement}. At each time step $t$, $t \in \{0, 1,\dots,T\}$, a drone can travel from one trajectory point $N_{start}$ to another trajectory point $N_{end}$ on the same square. The traveling distance of a drone at time step $t$ is $d(N_{start}, N_{end})$. The maximum distance, $d_{max}$, a drone can travel in a time step is from the current trajectory point to the diagonal trajectory points on the square. If $d(N_{start}, N_{end}) < d_{max}$, the drone will hover at $N_{end}$ until the end of the time step. Fig.~\ref{fig:gridPattern} illustrates an example with nine trajectory points. A drone at the trajectory point $N_4$ can travel to any one of the eight adjacent trajectory points in one time step, including the four diagonal trajectory points, $N_0$, $N_2$, $N_6$, and $N_8$, and four non-diagonal trajectory points, $N_1$, $N_3$, $N_5$, and $N_7$. When the drone travels to any one of $\{N_1, N_3, N_5, N_7\}$, it will spend the rest of the time step hovering at the endpoint.  

The action space $\mathcal{A}$ consists of ten possible actions, which can be categorized into three groups. The first group contains eight actions which, in time step $t$, allow the drone to travel from one trajectory point $N_i$ to any one of the eight trajectory points $N_f$ that fulfill $d(N_i, N_f) \leq d_{max}$. The first group of actions updates the trajectory point of a drone and consumes the power accordingly. The second group contains one action which allows a drone to hover at the current trajectory point. If the current trajectory point is the base station $N_0$, the drone is considered to be stationary and does not consume energy. Otherwise, the drone consumes energy to hover at the current point and maintain its state. The third group contains one action which allows a drone to execute the task at its current trajectory point. This action does not update the trajectory point of a drone but consumes the power of the drone for both task execution and hovering. It also changes the remaining length of the task if the drone is located at the task point. At the beginning of each time step $t$, each drone $k$ chooses an action $a \in \mathcal{A}$ and updates its location, battery level $B_k$, and the remaining task length accordingly.

The state space $\mathcal{S}$ is five-dimensional. For each time step $t$, it is denoted by Equation~\ref{eq:statespace}, 

\begin{equation}
S_t=\langle L_{task}^{\mathcal{K}}, L_{drone,t}^{\mathcal{K}}, A_{t}^{\mathcal{K}}, \tau_{t}^{\mathcal{K}}, B_{t}^{\mathcal{K}} \rangle,
\label{eq:statespace}
\end{equation} 

\noindent where $ L_{task}^{\mathcal{K}}$ represents all $K$ trajectory points where tasks are located, $L_{drone,t}^{\mathcal{K}}$ represents all $K$ trajectory points where drones are located at time step $t$, $A_{t}^{\mathcal{K}}$ represents all the actions taken by $K$ drones at time step $t$, $\tau_{t}^{\mathcal{K}}$ represents the remaining time steps to complete each of the $K$ tasks at current time step $t$, and $B_{t}^{\mathcal{K}}$ represents the battery levels of all $K$ drones at current time step $t$. Each of the notations can be defined by the following Equations~\ref{eq:L_task_k constraint} to~\ref{eq:B_t_k constraint}. 

\begin{equation}
L_{task}^{\mathcal{K}}  = \{l_{task}^k | l_{task}^k \in \mathcal{N} \backslash \{N_0\}, \forall k \in \mathcal{K}\},
\label{eq:L_task_k constraint}
\end{equation} 

\begin{equation}
L_{drone,t}^{\mathcal{K}}  = \{l_{drone,t}^{k} | l_{drone,t}^{k} \in \mathcal{N}, \forall k \in \mathcal{K}\},
\label{eq:L_drone,t_k constraint}
\end{equation} 
 
\begin{equation}
A_{t}^{\mathcal{K}} = \{a_{t}^{k} | a_{t}^{k} \in \mathcal{A}, \forall k \in \mathcal{K}\},
\label{eq:A_t_k constraint}
\end{equation} 

\begin{equation}
\tau_{t}^{\mathcal{K}} = \{\tau_{t}^{k} | 0 \leq \tau_{t}^{k} \leq \mathcal{T}_k, \forall k \in \mathcal{K}\},
\label{eq:tau_t_k constraint}
\end{equation} 

\begin{equation}
B_{t}^{\mathcal{K}} = \{B_{t}^{k} | B_{t}^{k} \leq B, \forall k \in \mathcal{K}\}.
\label{eq:B_t_k constraint}
\end{equation}

\subsection{Reward and Control Policy}

In each time step $t$, each drone chooses an action $a_t$ that transfers its state from $S_t$ to $S_{t+1}$ and receives an immediate reward $R_{t+1}$. The energy-aware control policy is then formulated by optimizing the accumulated reward. As our study concerns the battery capacity of each drone and the completion of the mission, the reward is formulated based on the execution progress of all tasks and the remaining energy of the drones. The dependence on the execution progress of all the tasks also enforce the collaboration among the agents to finish the tasks. To explore the impact of credit assignment in MARL, we compare our proposed individual reward mode with the shared reward mode proposed in~\cite{Li2022MARL}. We refer readers to the previous paper for the detailed formulation of the shared reward mode.

The individual reward concerns the state space change in time step $t$ associated with the action of each drone. Each drone receives a reward calculated based on its own action and the impact of the action on the global state. To simplify the process, we consider that each time step $t$ can be divided into a finite number of steps for drones to perform actions in order. The individual reward is calculated based on the state space change after a drone performs an action. Then, the system proceeds to the next drone. The task execution progress for each time step $t$, and each drone $k$ in individual reward simulation is defined by Equation~\ref{eq:taskexecution indv}, 

\begin{equation}
E(t, k) = \sum_{k=1}^{K}(\tau^k_{t-1}-\tau^k_{t}),
\label{eq:taskexecution indv}
\end{equation} 

\begin{algorithm}[ht]
\caption{Energy-Aware Multi-Agent DQN with experience replay for Collaborative Trajectory Planning and Task Execution}
\label{alg:strategyoutline}
\begin{algorithmic}
\FOR{Each drone $k = 1, \dots, K$}
  \STATE Initialize $\mathbb{E}$, a policy network parameters $\theta$, and a target network parameters $\theta^{-1}$ 
\ENDFOR
\FOR{Each episode}
    \STATE Initialize $S_0$
    \FOR{$t = 0, \dots ,T$} 
        \FOR{Each drone $k = 1, \dots, K$}
            \STATE Select action $A_t$ from $\mathcal{A}$ with given probability $\epsilon$
       	\STATE Generate a random probability 
            \IF { The probability $\leq \Delta$}
	   	   \STATE Choose an action from $\mathcal{A}$ via exploration
	    \ELSE
	   	   \STATE  Choose an action from $\mathcal{A}$ via exploitation
	    \ENDIF
            \STATE Execute the selected action and update the state to $S_{t+1}$
            \STATE observe the individual reward $R_{t+1, k}$
            \STATE Update experience replay $\mathbb{E}$ of drone $k$ with current sample $\{S_t, A^k_t, R_{t+1,k}, S_{t+1}\}$
            \STATE Pass $batch\_size \times \psi$ of samples from the replay memory of drone $k$ to the policy network
            \STATE Adopt Adam optimizer to train DQN using loss function \ref{eq:loss}
            \STATE Updates $\theta$ and $Q(s,a;\theta)$
            \STATE Update $\theta^{-1}$ with $\theta$ every $f$ time steps.
        \ENDFOR
    \ENDFOR
\ENDFOR
\end{algorithmic}
\end{algorithm}

The reward function is then defined by Equation~\ref{eq: individual reward},

\begin{equation}
\resizebox{0.48\textwidth}{!}{$
    R_{t+1, k}= 
\begin{cases}
    E(t+1, k) + \alpha \mu_{t+1,k},              & \text{if formulae~\ref{eq:constraint1} and~\ref{eq:constraint2} are true,}\\
    E(t+1, k) - \beta,              & \text{if formula~\ref{eq:constraint1} is true and formula~\ref{eq:constraint2} is false}, \\
    E(t+1, k),                                             & \text{if formula~\ref{eq:constraint1} is false,}
\end{cases}
$}
\label{eq: individual reward}
\end{equation} 

\noindent where, $\alpha$ and $\beta$ are two coefficients that adjust the importance of $\mu_{t+1, k}$ and the penalty for drone k running out of battery to achieve better energy efficiency, $\mu_{t+1, k}$ represents the percentage of the amount of remaining energy to the initial energy of drone $k$ at time step $t+1$.  $\mu_{t,k}$ is defined by Equation~\ref{eq:energyleft_ind},

\begin{equation}
\mu_{t,k} = \frac{B^k_t}{B}, 
\label{eq:energyleft_ind}
\end{equation} 

The reward function determines the optimal criterion. At each state, an action is associated with an expected accumulated reward $R(s,a)$ which represents the expected accumulated reward of an agent taking action $a$ at state $s$. A value $Q(s,a)$ then can be derived from $R(s,a)$ for each state-action pair and the Q function is given by the Bellman equation:

\begin{equation}
Q(s, a) = R(s,a) + \gamma \sum_{s'\in S}P(s'|s,a)\max_{a'\in A}Q(s',a'), 
\label{eq:Q_function}
\end{equation} 

\noindent Here $P(s'|s,a)$ is the transition probability of the states in the environment where $s'=s_{t+1}$, $\gamma$ is the trade-off between the importance of immediate and future rewards and it has the value $[0,1]$. The optimal policy is thus defined as Equation~\ref{eq:policy}:

\begin{equation}
\pi(s) = \text{arg}\max_{a\in A}Q(s,a), 
\label{eq:policy}
\end{equation} 

DQN forms an optimal policy by estimating Q value with a deep neural network which has the loss function as Equation~\ref{eq:loss} where $R$ denotes $R(s,a)$ for simplicity:
\begin{align}
    L(\theta) = \mathbb{E}_{(s,a,r,s')}[(R+\gamma\max_{a'}Q(s',a';\theta^{-1}) -Q(s,a;\theta))^2], 
    \label{eq:loss}
\end{align}

\noindent where $\mathbb{E}_{(s,a,r,s')}$ reflects the experience replay model that agent randomly samples past experience for learning, and $\theta$ refers to the network weights obtained by training.

In Algorithm~\ref{alg:strategyoutline}, we describe the training process of our proposed model. Each drone is initialized with a target network and a policy network. For each time step of each episode, each drone selects an action based on the \emph{exploration rate} $\Delta$ and the random probability $\epsilon$. If $\Delta > \epsilon$, the drone will randomly select an action. Otherwise, the drone selects an action with the highest Q-value at the present state. Depending on the reward mode, the action either results in a shared reward or an individual reward which is then saved to the experience replay model $\mathbb{E}$ coupled with the current state, action, and the next state. As the samples in $\mathbb{E}$ reach the threshold $batch\_size \times \psi$, the model will start to learn by sampling from past experience. The target network is updated every $f$ time step. The value for the three parameters $\psi$, $\Delta$, and $f$ will be explored in the next section.

\section{Performance Evaluation}
\label{performance eval}
\begin{figure}[htbp]
\centerline{\includegraphics[scale=0.35]{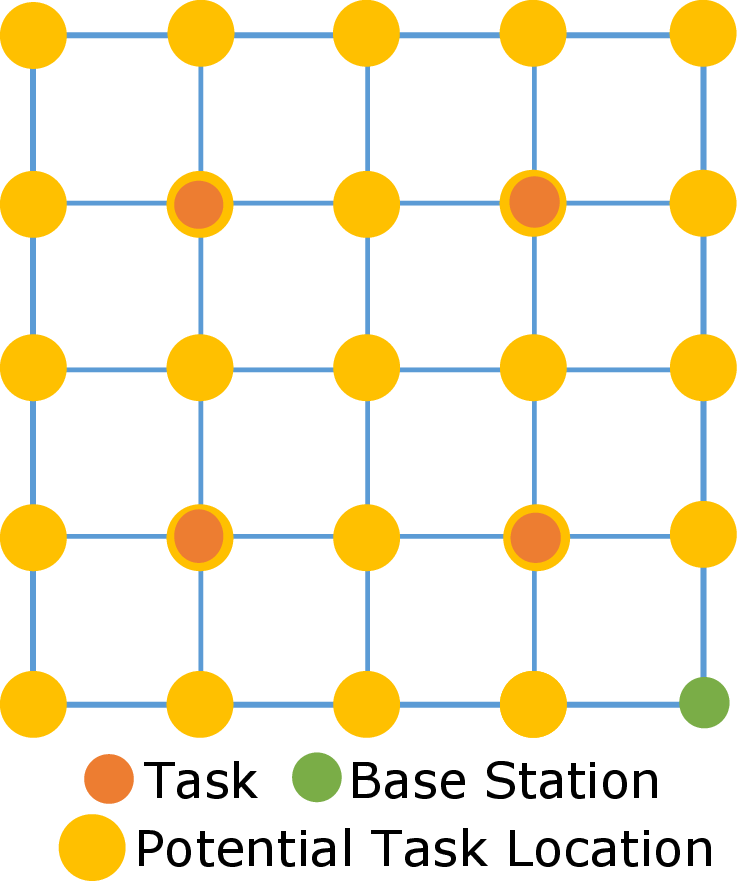}}
\caption{A $5 \times 5$ grid pattern consisting of 25 trajectory points with four tasks and one base station.}
\label{fig:4 tasks}
\end{figure}

We start our simulation study in a $5 \times 5$ grid, consisting of 25 trajectory points as shown in Fig.~\ref{fig:4 tasks}. With the adopted energy model proposed in~\cite{Stolaroff_nature_2018}, we obtained a set of energy consumption rates for our simulation shown in Table~\ref{tab:parameter}.

\begin{table}[htbp]
\caption{Base Case Parameters to Calculate Energy Consumption Rates}
\label{tab:parameter}
\centerline{\begin{tabular}{cccccc}%
\hline
 \textit{$D$[m]} & \textit{$m_{drone}$[kg]} & \textit{$v$[m/s]}  & \textit{$\alpha$[rad]} & \textit{$\eta$} & \textit{$\rho$[$kg/m^2$]}    \\  [0.5ex]
\hline
\hline
 0.254  & 2.07  & 10 & 0.0139  & 0.7 &  1.2193\\ [0.5ex]
\hline
\end{tabular}}
\end{table}

To simplify the calculation for easy understanding and focus on the performance comparison in our experiment, we proportionally scale down the energy consumption rate values based on the calculated values to $P_{E} = 3$, $P_{H} = 4$, and $P_{F} = 2.5$. We set the maximum time step per episode as $t = 600$. If the tasks are not finished within the maximum time step, the episode will terminate with a failure indicator. Considering the number of time steps for each episode in our simulation, we set the battery capacity of each drone $B = 1800$ for both shared reward mode and individual reward mode to ensure sufficient energy for exploration and learning task execution strategies to minimize energy consumption. We note that the values of energy consumption rate and battery capacity can be proportionally individualized for specific scenarios.

We explored four sets of experiments in this paper. The first set is to understand the impact of batch size $\psi$, target network update frequency $f$, and the threshold of exploration rate $\Delta$. Through the analysis of the three parameters, we find a set of parameters that generates the best performance to explore the second experiment, which is to show the robustness of the model facing non-predetermined task lengths and locations. The third set is to inspect the influence of task density on the performance of our proposed model. Lastly, we compare the agents' performance under the individual reward scenario with the shared reward to explore the reward assignment impact and shows the advantages of our proposed individual reward mode. To form a standard comparison, we use \emph{success rate} and \emph{average execution steps} to measure the performance of the proposed model. The success rate is calculated as the ratio of the number of accomplished missions to the total number of missions within an exploration period. The average execution steps are calculated as the average number of steps required to finish a mission within an exploration period.

\subsection{Parameter Study}

We explore $\psi$, $f$, and cut off $\Delta$ the three parameters in this set of experiments. When one parameter is evaluated, the other two remain unchanged. The simulation environment is controlled with $K=4$ at fixed locations and length $\mathcal{T}_i = 5, \forall i \in \mathcal{K}$. The four task locations are shown as Fig.~\ref{fig:4 tasks}. The initial $\delta$ is set to 0.5 and gradually decreased by $3 \times 10^{-6}$ per time step. 

\begin{figure}[h]
    \centerline{\includegraphics[scale=0.45]{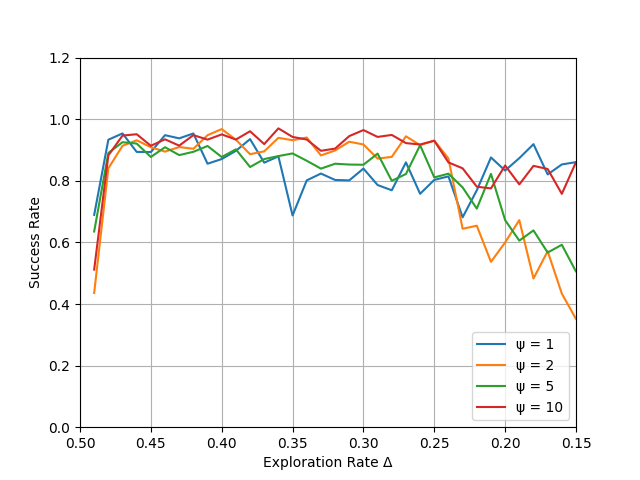}}
    \caption{The success rate under different batch size $\psi$ values and exploration rates.}
    \label{fig 4 a:batch}
\end{figure}

\begin{figure}[h]
    \centerline{\includegraphics[scale=0.45]{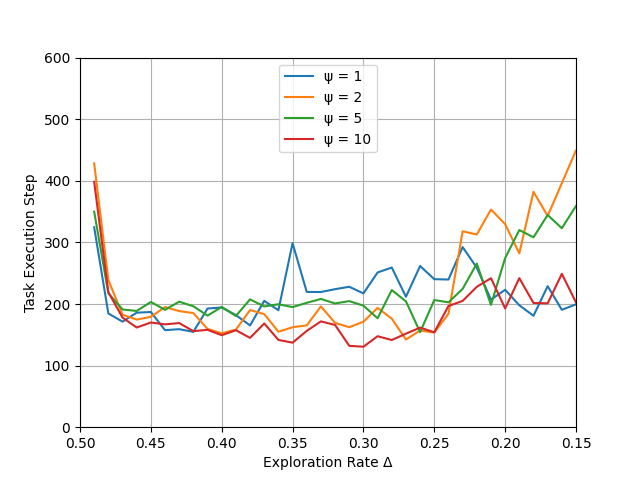}}
    \caption{The average number of steps required to finish a mission under different batch size $\psi$ values and exploration rates.}
    \label{fig 4 b:batch}
\end{figure}

We firstly evaluate the value for $\psi$ following the set of values \{1, 2, 5, 10\}. The results for the evaluation of $\psi$ are shown in Fig.~\ref{fig 4 a:batch} and Fig.~\ref{fig 4 b:batch}. Fig.~\ref{fig 4 a:batch} shows the success rate steadily increases with all choices of batch sizes, with a fluctuation at the end. Fig.~\ref{fig 4 b:batch} shows the average task execution steps also decrease with a fluctuation approaching the end of training which differentiates the choices. $\psi = 1$ and $\psi = 10$ stand out compared to other values with higher success rates and less average task execution steps. However, we chose $\psi = 10$ as the final value as it results in a more stable performance in both success rate analysis and task execution step analysis. Based on these two figures, we also decided the cut-off value for $\Delta$ to be 0.15 to maintain high performance while enabling the agents to use the learned experiences mostly. 

Note that in the above analysis, both plots reveal the same trend. But the evaluations based on task execution steps show the task execution efficiency and amplify the differences among different parameter values during the learning process. For better clarity of the analysis, we include the plots of evaluations based on task execution steps for the rest of the parameter studies. The plots of evaluations based on the success rate for the rest of the parameters studies are included in Appendix.

With $\psi = 10$ and cut-off value for $\Delta = 0.15$, we explored the value for $f$ from the following series \{500, 1000, 2000, 4000, 8000\}. The results for the evaluation of $f$ are shown in Fig.~\ref{fig 5:freq}, and the evaluations based on success rate are included in Appendix~\ref{app A: parameter}. All of the values give similar performance according to the graph. We chose $f = 8000$ as the most suitable value as it shows a more stable and better performance approaching the end of training. 

\begin{figure}[h]
    \centerline{\includegraphics[scale=0.45]{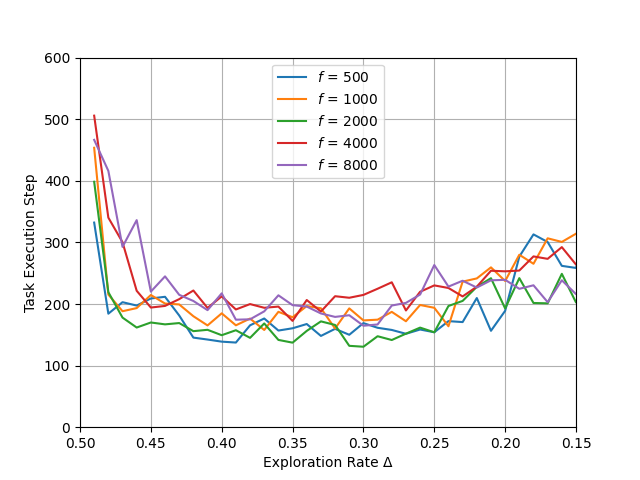}}
    \caption{The average number of steps required to finish a mission under different target network update frequency \textit{f} values and exploration rates.}
    \label{fig 5:freq}
\end{figure}

\subsection{Task Length and Location Study}

\begin{figure}[h]
    \centerline{\includegraphics[scale=0.45]{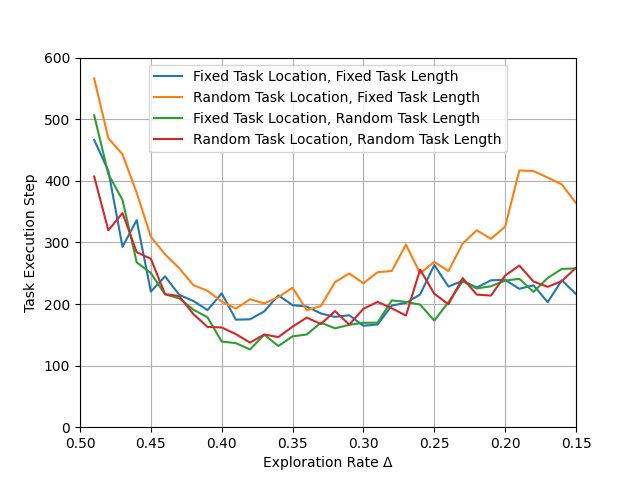}}
    \caption{The average number of steps required to finish a mission under different task locations and task length conditions.}
    \label{fig 6:random}
\end{figure}

With the selected values for $\psi$, $f$, and cut-off $\Delta$, we evaluated the proposed model with another three different scenarios, including random task locations with fixed task lengths, fixed task locations with random task lengths, and random task locations with random task lengths. When the task locations are fixed, they are located at trajectory points indicated by Fig.~\ref{fig:4 tasks}. As the location changes, they can be any four points on the grid except the base station. When the task lengths are fixed, $\mathcal{T}_i = 5, \forall i \in \mathcal{K}$. As the task length changes, a task length has an integer value in the range $[1, 5]$. The results of the three scenarios are plotted in Fig.~\ref{fig 6:random} with the additional scenario where both task locations and task lengths are fixed. Same reason as the one in the parameter study section, we include the evaluations based on success rate in Appendix~\ref{app B: random} for your reference.

Except for the scenario of random task locations with fixed task lengths, the required task execution steps are decreased significantly for all other three scenarios. This indicates that our proposed model enables drones to accomplish missions in dynamic environments under various unpredictable changes. It is noted that the scenarios with random task lengths have a better performance due to the shorter mission lengths on average. The graph also shows that the performance of the model in the scenario of random task location with fixed task length is worse than other cases. This is caused by the fact that the task lengths remain at the highest value while the task location changes, which proposes greater difficulty in learning. This scenario may require to lower value for exploration decrement rate and longer training.  

\subsection{Task Density Study}

\begin{figure}[h]
    \centerline{\includegraphics[scale=0.45]{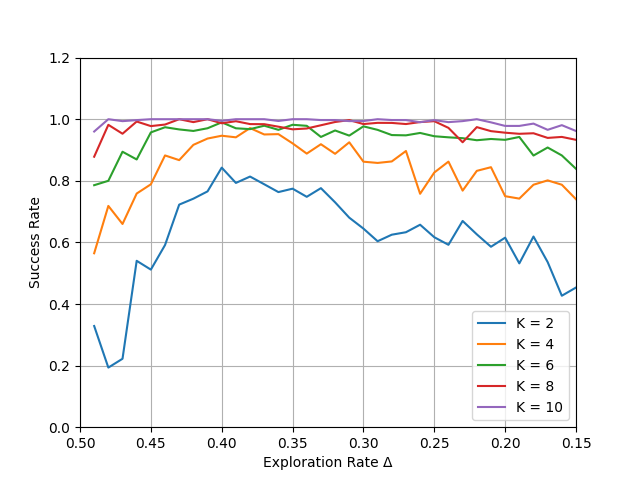}}
    \caption{The success rate under different numbers of tasks K values and exploration rates.}
    \label{fig 7 a:density}
\end{figure}

\begin{figure}[h]
    \centerline{\includegraphics[scale=0.45]{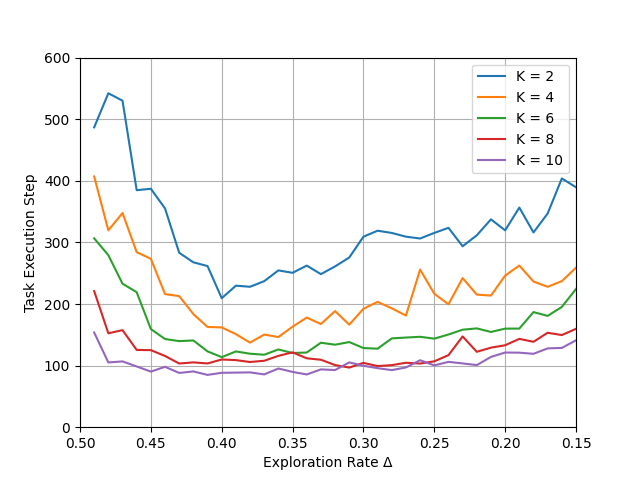}}
    \caption{The average number of steps required to finish a mission under different numbers of tasks K values and exploration rates.}
    \label{fig 7 b:density}
\end{figure}

In this study, we explore task density's impact on our proposed model's performance. Task density here refers to the ratio of the number of tasks to the total number of trajectory points. The task numbers follow the series \{2, 4, 6, 8, 10\}. Throughout this study, for each episode, the tasks are randomly located on the 25 trajectory points except the base station, and the task lengths are integers that fall randomly within the range $[1, 5]$. The results of this set of experiments based on task execution steps are plotted in Fig.~\ref{fig 7 a:density} and Fig.~\ref{fig 7 b:density}.%

As Fig.~\ref{fig 7 a:density} and Fig.~\ref{fig 7 b:density} indicates, task density affects the performance of the model positively. As the task density increases, the required task execution steps decrease, shown in the Fig.~\ref{fig 7 b:density}. When task number $K = 10$, the proposed model achieved nearly 100\% success rate, as shown in Fig.~\ref{fig 7 a:density}. This is expected when the task density is high, and it is more likely for the drones to arrive at the task locations and execute tasks during exploration, which enables faster learning.

\subsection{Grid Size Study}
To further examine the robustness of the proposed individual-reward model, we conducted this set of experiments to evaluate the performance of the models across different grid sizes. The width of the grid follows the series $\{5, 6, 7, 8\}$ and the shape follows the illustration in Fig.~\ref{fig:4 tasks}. To maintain the same task density for all experiments in this set, we consider the scenario of $K = 10$ for grid size $5 \times 5$ as the standard since it gives the best performance among the experiments conducted in the task density study. This implies that we set $K = 15$ for grid size $6 \times 6$, $K = 20$ for grid size $7 \times 7$, and $K= 26$ for grid size $8 \times 8$. The values of hyperparameters $\psi$, $f$, and $\Delta$ are kept the same as before, and the number of agents equals the number of tasks in each case. Throughout this set of experiments, the tasks are randomly located on the trajectory points except for the base station, and the task lengths are integers that fall randomly within the range [1, 5]. 

\begin{figure}[h]
    \centerline{\includegraphics[scale=0.45]{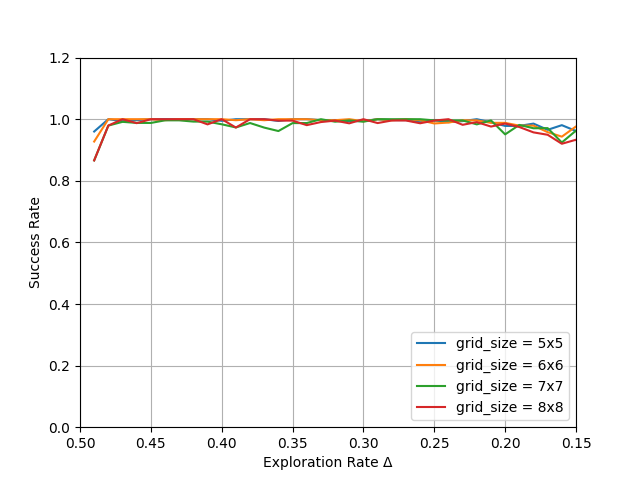}}
    \caption{The success rate under different grid sizes and exploration rates.}
    \label{fig 8 a:gridsize}
\end{figure}

\begin{figure}[h]
    \centerline{\includegraphics[scale=0.45]{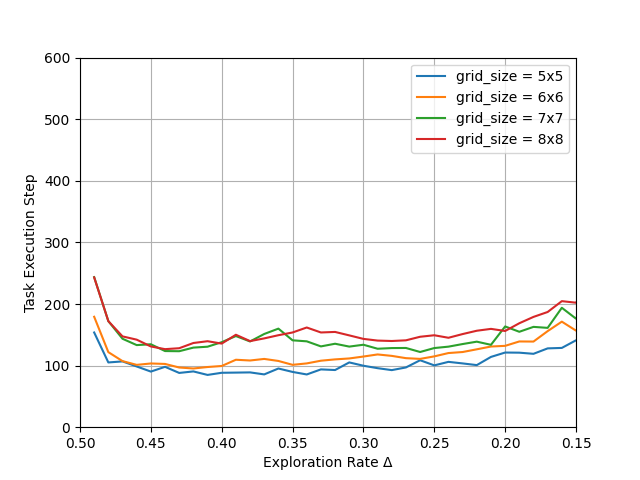}}
    \caption{The average number of steps required to finish a mission under different grid sizes and exploration rates.}
    \label{fig 8 b:gridsize}
\end{figure}

The evaluation results on success rate are displayed in Fig.~\ref{fig 8 a:gridsize}, and the results on the task execution steps are shown in Fig.~\ref{fig 8 b:gridsize}. While as the grid size increases, the average number of steps required to finish a mission increases slightly, and the success rates are quite stable and close to $100\%$. This is expected since a larger grid size requires the drones to take more steps for traveling in order to reach the tasks that are located farther from the base station. Even with the slight increase in the number of steps, the agents still show the capability to learn and execute the tasks efficiently with our proposed model. 

\subsection{Individual Reward vs. Shared Reward}

We explored the impact of reward assignments in this set of experiments. We conducted the comparison in both the task density study and scalability study. For both sets of comparison studies, we kept the exploration cut-off value of $\Delta$ as $0.15$, the same as the previous experiments, and evaluated the performance at this exploration rate which is the value at the end of the training. The required task execution steps and the success rate are used as the evaluation metric. Our newly proposed individual reward mode reveals its advantages when compared with the shared reward mode. We include the evaluation results here, and the plots of training are included in Appendix~\ref{app E: comp}, which shows the different convergence behavior of the individual reward mode and shared reward mode. The performance of the shared reward mode is largely affected by the increase of the grid size and fails to converge when the grid size increases to a certain level. Whereas, our proposed individual reward mode is more robust towards the increase of the grid size and manages to obtain a good performance across different grid sizes.

\begin{figure}[h]
    \centerline{\includegraphics[scale=0.45]{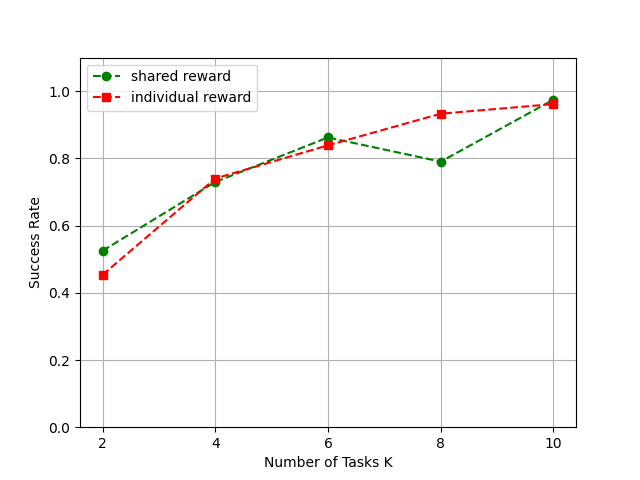}}
    \caption{Performance comparison on success rate under different task densities between the individual and shared rewards modes.}
    \label{fig 9 a:comparison}
\end{figure}

\begin{figure}[h]
    \centerline{\includegraphics[scale=0.45]{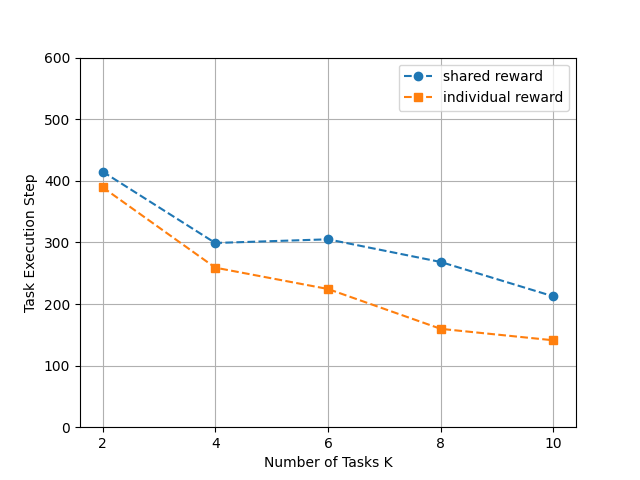}}
    \caption{Performance comparison on task execution steps under different task densities between the individual and shared rewards modes. }
    \label{fig 9 b:comparison}
\end{figure}

\begin{figure}[h]
    \centerline{\includegraphics[scale=0.45]{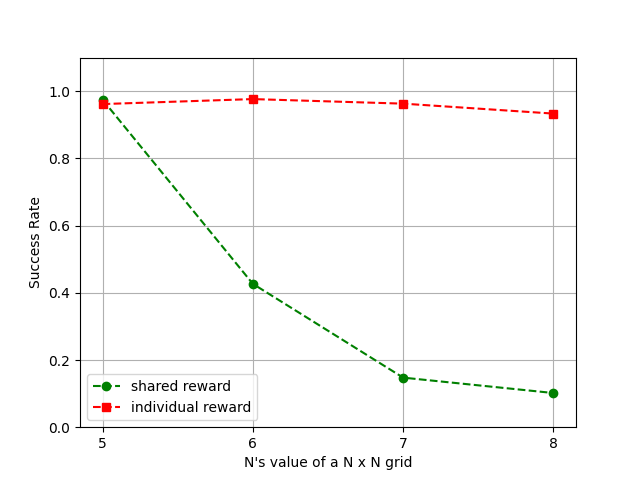}}
    \caption{Performance comparison on success rate under different grid sizes between the individual and shared rewards modes.}
    \label{fig 10 a:comparison}
\end{figure}

\begin{figure}[h]
    \centerline{\includegraphics[scale=0.45]{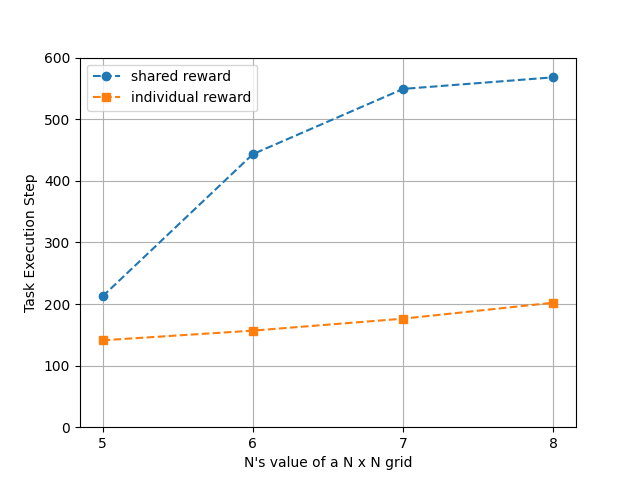}}
    \caption{Performance comparison on task execution steps under different grid sizes between the individual and shared rewards modes. }
    \label{fig 10 b:comparison}
\end{figure}

The comparison of task density is shown in Fig.~\ref{fig 9 a:comparison} and Fig.~\ref{fig 9 b:comparison}. We compared the performance over task numbers following the series \{2, 4, 6, 8, 10\} while keeping the grid size as $5\times5$. Both reward modes show a similar pattern: as the task density increases, the success rate approaches 100\%. They also share similar success rate values in all five scenarios, with individual reward mode showing more stable performance. As to the required task execution steps, both reward modes decrease while the task density increases. However, the individual reward mode performs better under all five different task densities. We note that the performance of individual reward mode is significantly better for scenarios of 6 tasks and 8 tasks while the performance improvement becomes less significant for 10 tasks. This phenomenon can be attributed to the opposing effects of reward assignment and the difficulty of obtaining reward signals. In environments with high task density, agents can acquire reward signals more readily by reaching task locations more frequently. This increased accessibility reduces the complexity of learning policies compared to environments with lower task density, such as those with 6 or 8 tasks. Consequently, the ease of obtaining reward signals diminishes the impact of reward assignment in shared reward settings, resulting in a smaller performance gap between individual reward modes and shared reward modes.

The effect of the credit assignment problem becomes more apparent in the comparison for scalability evaluation. We vary the grid size from $5\times 5$ to $8 \times 8$. The evaluation results are shown in Fig.~\ref{fig 10 a:comparison} and Fig.~\ref{fig 10 b:comparison}. The success rate of the shared reward mode drops drastically as the grid size increases. The required task execution step also increases to a level close to the cut-off time step, indicating the failure to learn the optimal policy when the grid size increases. However, our proposed individual reward mode maintains a high success rate close to $100\%$ across different grid sizes, and the required task execution step increases slowly as the grid size increases, indicating that the performance is only mildly impacted by the grid size. This performance gap can be attributed to reward assignment problem in shared reward. As there are more agents, the reward signal for whether an individual agent contributing to the global goal becomes unclear in shared reward mode. Thus, the agents falsely update their Q-value and assign the incorrect values to the state-action pairs due to the obscure signals received from the reward. While in individual reward mode, the agents receive rewards based on their individual actions. Thus, each agent has a clear goal, and the learning process is less ambiguous, which results in faster learning and better performance. Both task density and grid size evaluations prove that the individual reward mode is more robust and scalable than the shared reward mode.

\section{Conclusion and Future Work}
\label{conclusion}

In this paper, we propose an energy-aware task execution MARL model for mission-oriented drone networks to combat the challenge of the limited battery capacity of drones using an individual reward function. Each drone has its own DQN, which is trained to guide the trajectory planning, task assignment, and task execution of the drones. The task execution progress and the remaining battery levels drive the individual reward function. 

To evaluate our proposed model, we create a simulation environment that allows flexible changes of task lengths and locations. We focused on evaluating the individual reward mode and comparing the individual reward mode and shared reward mode. We conducted three sets of experiments on the individual reward mode to investigate the impact of the hyperparameters, including $\psi$, $f$ and cut-off $\Delta$, the dynamic environment, and the task density. The evaluation results show that our proposed model can accomplish the mission with a success rate of 80\% for most cases in a dynamic and unpredictable environment after training. The model was even able to achieve nearly 100\% success rate for scenarios where the task density is relatively high. The comparison of performance between the two modes under different task densities and grid sizes shows that the individual reward mode is more robust to the change of the environment size and task density as it achieves better performance consistently, which can be explained by the more clear reward assignment of MARL.  

However, the evaluation of our proposed model also reveals several limitations that require future work. Firstly, the performance is unstable in some instances, such as the scenario of random task locations with fixed task lengths. The performance also fluctuates when approaching the end of the training, which might indicate that the exploration decrement needs to be lower to allow agents to have more exploration. In addition, the outliers in the distribution of experiences can also largely influence performance, which is a common issue with reinforcement learning and requires further study to investigate how to discount those bad experiences and ensure a more stable performance. Some recent works have proposed different new algorithms to stabilize the performance of MARL such as in~\cite{li2019robust}, \cite{rashid2020monotonic}, \cite{son2019qtran} and \cite{yang2018mean}. We note that these studies fundamentally differ from ours. They focus on improving MARL algorithm itself but we create a practical framework specifically for an application scenario while maintaining the simplicity of MARL structure for embedding consideration. However, future work can explore whether those proposed algorithms are applicable in drone networks and bring further improvement in performance with the consideration of loading neural networks in embedded systems.

On the other hand, our current simulation concerns the 2D environment. We will extend the environment to 3D and evaluate the model's performance in the 3D environment in our future study, which should ensure better applicability of our proposed model. What's more, additional evaluation of our proposed method applied to real mission-oriented drone networks for cooperatively executing real-world tasks can be explored to showcase the advantages. Technical challenges such as collision avoidance and information exchange need to be considered to make it more practical and applicable which can be solved by combining with techniques proposed in~\cite{shen2022multidepot}, \cite{yanmaz2018drone}. However, this is not within the scope of this study.

\section*{Acknowledgments}
We thank Colby College for the compute support.

\bibliographystyle{IEEEtran}
\bibliography{IEEEexample}

\begin{thebibliography}{10}
\providecommand{\url}[1]{#1}
\csname url@samestyle\endcsname
\providecommand{\newblock}{\relax}
\providecommand{\bibinfo}[2]{#2}
\providecommand{\BIBentrySTDinterwordspacing}{\spaceskip=0pt\relax}
\providecommand{\BIBentryALTinterwordstretchfactor}{4}
\providecommand{\BIBentryALTinterwordspacing}{\spaceskip=\fontdimen2\font plus
\BIBentryALTinterwordstretchfactor\fontdimen3\font minus \fontdimen4\font\relax}
\providecommand{\BIBforeignlanguage}[2]{{%
\expandafter\ifx\csname l@#1\endcsname\relax
\typeout{** WARNING: IEEEtran.bst: No hyphenation pattern has been}%
\typeout{** loaded for the language `#1'. Using the pattern for}%
\typeout{** the default language instead.}%
\else
\language=\csname l@#1\endcsname
\fi
#2}}
\providecommand{\BIBdecl}{\relax}
\BIBdecl

\bibitem{Li2022MARL}
Y.~Li, C.~Li, J.~Chen, and C.~Roinou, ``{Energy-Aware Multi-Agent Reinforcement Learning for Collaborative Execution in Mission-Oriented Drone Networks},'' \emph{IEEE International Conference on Computer Communications and Networks}, July. 2022.

\bibitem{mnih2015human}
V.~Mnih, K.~Kavukcuoglu, D.~Silver, A.~A. Rusu, J.~Veness, M.~G. Bellemare, A.~Graves, M.~Riedmiller, A.~K. Fidjeland, G.~Ostrovski \emph{et~al.}, ``{Human-level control through deep reinforcement learning},'' \emph{nature}, vol. 518, no. 7540, pp. 529--533, 2015.

\bibitem{hong2017deep}
Z.-W. Hong, S.-Y. Su, T.-Y. Shann, Y.-H. Chang, and C.-Y. Lee, ``A deep policy inference q-network for multi-agent systems,'' \emph{arXiv preprint arXiv:1712.07893}, 2017.

\bibitem{oroojlooyjadid2019review}
A.~OroojlooyJadid and D.~Hajinezhad, ``{A review of cooperative multi-agent deep reinforcement learning},'' \emph{arXiv preprint arXiv:1908.03963}, 2019.

\bibitem{yao2019collaborative}
F.~Yao and L.~Jia, ``A collaborative multi-agent reinforcement learning anti-jamming algorithm in wireless networks,'' \emph{IEEE wireless communications letters}, vol.~8, no.~4, pp. 1024--1027, 2019.

\bibitem{tampuu2017multiagent}
A.~Tampuu, T.~Matiisen, D.~Kodelja, I.~Kuzovkin, K.~Korjus, J.~Aru, J.~Aru, and R.~Vicente, ``Multiagent cooperation and competition with deep reinforcement learning,'' \emph{PloS one}, vol.~12, no.~4, p. e0172395, 2017.

\bibitem{bucsoniu2010multi}
L.~Bu{\c{s}}oniu, R.~Babu{\v{s}}ka, and B.~De~Schutter, ``Multi-agent reinforcement learning: An overview,'' \emph{Innovations in multi-agent systems and applications-1}, pp. 183--221, 2010.

\bibitem{lowe2017multi}
R.~Lowe, Y.~I. Wu, A.~Tamar, J.~Harb, O.~Pieter~Abbeel, and I.~Mordatch, ``Multi-agent actor-critic for mixed cooperative-competitive environments,'' \emph{Advances in neural information processing systems}, vol.~30, 2017.

\bibitem{chen2020delay}
B.~Chen, M.~Xu, Z.~Liu, L.~Li, and D.~Zhao, ``Delay-aware multi-agent reinforcement learning for cooperative and competitive environments,'' \emph{arXiv preprint arXiv:2005.05441}, 2020.

\bibitem{schaul2015prioritized}
T.~Schaul, J.~Quan, I.~Antonoglou, and D.~Silver, ``Prioritized experience replay,'' \emph{arXiv preprint arXiv:1511.05952}, 2015.

\bibitem{haarnoja2018soft}
T.~Haarnoja, A.~Zhou, P.~Abbeel, and S.~Levine, ``Soft actor-critic: Off-policy maximum entropy deep reinforcement learning with a stochastic actor,'' in \emph{International conference on machine learning}.\hskip 1em plus 0.5em minus 0.4em\relax PMLR, 2018, pp. 1861--1870.

\bibitem{fujimoto2018addressing}
S.~Fujimoto, H.~Hoof, and D.~Meger, ``Addressing function approximation error in actor-critic methods,'' in \emph{International conference on machine learning}.\hskip 1em plus 0.5em minus 0.4em\relax PMLR, 2018, pp. 1587--1596.

\bibitem{li2024roer}
C.~Li, Z.-W. Hong, P.~Agrawal, D.~Garg, and J.~Pajarinen, ``Roer: Regularized optimal experience replay,'' \emph{arXiv preprint arXiv:2407.03995}, 2024.

\bibitem{vazquez2019multi}
J.~Vazquez-Canteli, T.~Detjeen, G.~Henze, J.~K{\"a}mpf, and Z.~Nagy, ``{Multi-agent reinforcement learning for adaptive demand response in smart cities},'' in \emph{Journal of Physics: Conference Series}, vol. 1343, no.~1.\hskip 1em plus 0.5em minus 0.4em\relax IOP Publishing, 2019, p. 012058.

\bibitem{duan2012multi}
Y.~Duan, B.~X. Cui, and X.~H. Xu, ``{A multi-agent reinforcement learning approach to robot soccer},'' \emph{Artificial Intelligence Review}, vol.~38, no.~3, pp. 193--211, 2012.

\bibitem{zhang2019integrating}
Z.~Zhang, J.~Yang, and H.~Zha, ``{Integrating independent and centralized multi-agent reinforcement learning for traffic signal network optimization},'' \emph{arXiv preprint arXiv:1909.10651}, 2019.

\bibitem{9537638}
Z.~Zhang, H.~Liu, M.~Zhou, and J.~Wang, ``Solving dynamic traveling salesman problems with deep reinforcement learning,'' \emph{IEEE Transactions on Neural Networks and Learning Systems}, vol.~34, no.~4, pp. 2119--2132, 2023.

\bibitem{10153615}
C.~Wang, Z.~Bi, and Y.~Wan, ``Secure underwater distributed antenna systems: A multi-agent reinforcement learning approach,'' \emph{IEEE/CAA Journal of Automatica Sinica}, vol.~10, no.~7, pp. 1622--1624, 2023.

\bibitem{pugliese2020using}
L.~D.~P. Pugliese, F.~Guerriero, and G.~Macrina, ``{Using drones for parcels delivery process},'' \emph{Procedia Manufacturing}, vol.~42, pp. 488--497, 2020.

\bibitem{10606397}
J.~Zhang, Y.~Wu, and M.~Zhou, ``Cooperative dual-task path planning for persistent surveillance and emergency handling by multiple unmanned ground vehicles,'' \emph{IEEE Transactions on Intelligent Transportation Systems}, pp. 1--12, 2024.

\bibitem{10606273}
H.~Yuan, M.~Wang, J.~Bi, S.~Shi, J.~Yang, J.~Zhang, M.~Zhou, and R.~Buyya, ``Cost-efficient task offloading in mobile edge computing with layered unmanned aerial vehicles,'' \emph{IEEE Internet of Things Journal}, pp. 1--1, 2024.

\bibitem{10445189}
R.~Lai, B.~Zhang, G.~Gong, H.~Yuan, J.~Yang, J.~Zhang, and M.~Zhou, ``Energy-efficient scheduling in uav-assisted hierarchical wireless sensor networks,'' \emph{IEEE Internet of Things Journal}, vol.~11, no.~11, pp. 20\,194--20\,206, 2024.

\bibitem{10415630}
J.~Kang, J.~Chen, M.~Xu, Z.~Xiong, Y.~Jiao, L.~Han, D.~Niyato, Y.~Tong, and S.~Xie, ``Uav-assisted dynamic avatar task migration for vehicular metaverse services: A multi-agent deep reinforcement learning approach,'' \emph{IEEE/CAA Journal of Automatica Sinica}, vol.~11, no.~2, pp. 430--445, 2024.

\bibitem{hu2020cooperative}
J.~Hu, H.~Zhang, L.~Song, R.~Schober, and H.~V. Poor, ``{Cooperative internet of UAVs: Distributed trajectory design by multi-agent deep reinforcement learning},'' \emph{IEEE Transactions on Communications}, vol.~68, no.~11, pp. 6807--6821, 2020.

\bibitem{shihavuddin2019wind}
A.~Shihavuddin, X.~Chen, V.~Fedorov, A.~Nymark~Christensen, N.~Andre Brogaard~Riis, K.~Branner, A.~Bjorholm~Dahl, and R.~Reinhold~Paulsen, ``{Wind turbine surface damage detection by deep learning aided drone inspection analysis},'' \emph{Energies}, vol.~12, no.~4, p. 676, 2019.

\bibitem{toth2002vehicle}
P.~Toth and D.~Vigo, \emph{The vehicle routing problem}.\hskip 1em plus 0.5em minus 0.4em\relax SIAM, 2002.

\bibitem{hashemi2019new}
S.~R. Hashemi, R.~Esmaeeli, H.~Aliniagerdroudbari, M.~Alhadri, H.~Alshammari, A.~Mahajan, and S.~Farhad, ``{New intelligent battery management system for drones},'' in \emph{ASME international mechanical engineering congress and exposition}, vol. 59438.\hskip 1em plus 0.5em minus 0.4em\relax American Society of Mechanical Engineers, 2019, p. V006T06A028.

\bibitem{du2021survey}
W.~Du and S.~Ding, ``{A survey on multi-agent deep reinforcement learning: from the perspective of challenges and applications},'' \emph{Artificial Intelligence Review}, vol.~54, no.~5, pp. 3215--3238, 2021.

\bibitem{wang20213m}
W.~Wang, Y.~Liu, R.~Srikant, and L.~Ying, ``{3M-RL: Multi-Resolution, Multi-Agent, Mean-Field Reinforcement Learning for Autonomous UAV Routing},'' \emph{IEEE Transactions on Intelligent Transportation Systems}, 2021.

\bibitem{10309976}
H.~Xie, D.~Zhang, X.~Hu, M.~Zhou, and Z.~Cao, ``Autonomous multirobot navigation and cooperative mapping in partially unknown environments,'' \emph{IEEE Transactions on Instrumentation and Measurement}, vol.~72, pp. 1--12, 2023.

\bibitem{chen2021multi}
Y.-J. Chen, K.-M. Liao, M.-L. Ku, F.~P. Tso, and G.-Y. Chen, ``{Multi-Agent Reinforcement Learning Based 3D Trajectory Design in Aerial-Terrestrial Wireless Caching Networks},'' \emph{IEEE Transactions on Vehicular Technology}, vol.~70, no.~8, pp. 8201--8215, 2021.

\bibitem{zhao2020multi}
N.~Zhao, Z.~Liu, and Y.~Cheng, ``{Multi-agent deep reinforcement learning for trajectory design and power allocation in multi-UAV networks},'' \emph{IEEE Access}, vol.~8, pp. 139\,670--139\,679, 2020.

\bibitem{wang2020multi}
L.~Wang, K.~Wang, C.~Pan, W.~Xu, N.~Aslam, and L.~Hanzo, ``{Multi-agent deep reinforcement learning-based trajectory planning for multi-UAV assisted mobile edge computing},'' \emph{IEEE Transactions on Cognitive Communications and Networking}, vol.~7, no.~1, pp. 73--84, 2020.

\bibitem{khalil2021efficient}
A.~A. Khalil, A.~J. Byrne, M.~A. Rahman, and M.~H. Manshaei, ``{Efficient uav trajectory-planning using economic reinforcement learning},'' \emph{arXiv preprint arXiv:2103.02676}, 2021.

\bibitem{10183796}
G.~He, M.~Feng, Y.~Zhang, G.~Liu, Y.~Dai, and T.~Jiang, ``Deep reinforcement learning based task-oriented communication in multi-agent systems,'' \emph{IEEE Wireless Communications}, vol.~30, no.~3, pp. 112--119, 2023.

\bibitem{9980391}
D.~Liu, L.~Dou, R.~Zhang, X.~Zhang, and Q.~Zong, ``Multi-agent reinforcement learning-based coordinated dynamic task allocation for heterogenous uavs,'' \emph{IEEE Transactions on Vehicular Technology}, vol.~72, no.~4, pp. 4372--4383, 2023.

\bibitem{wang2020shapley}
J.~Wang, Y.~Zhang, T.-K. Kim, and Y.~Gu, ``{Shapley Q-value: A local reward approach to solve global reward games},'' in \emph{Proceedings of the AAAI Conference on Artificial Intelligence}, vol.~34, no.~05, 2020, pp. 7285--7292.

\bibitem{yang2018cm3}
J.~Yang, A.~Nakhaei, D.~Isele, K.~Fujimura, and H.~Zha, ``{Cm3: Cooperative multi-goal multi-stage multi-agent reinforcement learning},'' \emph{arXiv preprint arXiv:1809.05188}, 2018.

\bibitem{sheikh2020multi}
H.~U. Sheikh and L.~B{\"o}l{\"o}ni, ``{Multi-agent reinforcement learning for problems with combined individual and team reward},'' in \emph{2020 International Joint Conference on Neural Networks (IJCNN)}.\hskip 1em plus 0.5em minus 0.4em\relax IEEE, 2020, pp. 1--8.

\bibitem{du2019liir}
Y.~Du, L.~Han, M.~Fang, J.~Liu, T.~Dai, and D.~Tao, ``{Liir: Learning individual intrinsic reward in multi-agent reinforcement learning},'' \emph{Advances in Neural Information Processing Systems}, vol.~32, 2019.

\bibitem{Stolaroff_nature_2018}
J.~K. Stolaroff, C.~Samaras, E.~R. O’Neill, A.~Lubers, A.~S. Mitchell, and D.~Ceperley, ``Energy use and life cycle greenhouse gas emissions of drones for commercial package delivery,'' \emph{Nature communications}, vol.~9, no.~1, p. 409, 2018.

\bibitem{arnautovic2021evaluation}
A.~Arnautovi{\'c} and E.~Teskered{\v{z}}i{\'c}, ``Evaluation of artificial neural network inference speed and energy consumption on embedded systems,'' in \emph{2021 20th International Symposium INFOTEH-JAHORINA (INFOTEH)}.\hskip 1em plus 0.5em minus 0.4em\relax IEEE, 2021, pp. 1--5.

\bibitem{loni2019neuropower}
M.~Loni, A.~Zoljodi, S.~Sinaei, M.~Daneshtalab, and M.~Sj{\"o}din, ``Neuropower: Designing energy efficient convolutional neural network architecture for embedded systems,'' in \emph{Artificial Neural Networks and Machine Learning--ICANN 2019: Theoretical Neural Computation: 28th International Conference on Artificial Neural Networks, Munich, Germany, September 17--19, 2019, Proceedings, Part I 28}.\hskip 1em plus 0.5em minus 0.4em\relax Springer, 2019, pp. 208--222.

\bibitem{lee2019neuro}
S.~Lee and S.~Nirjon, ``Neuro. zero: a zero-energy neural network accelerator for embedded sensing and inference systems,'' in \emph{Proceedings of the 17th Conference on Embedded Networked Sensor Systems}, 2019, pp. 138--152.

\bibitem{kim2020energy}
B.~Kim, S.~Lee, A.~R. Trivedi, and W.~J. Song, ``Energy-efficient acceleration of deep neural networks on realtime-constrained embedded edge devices,'' \emph{IEEE Access}, vol.~8, pp. 216\,259--216\,270, 2020.

\bibitem{verhelst2017embedded}
M.~Verhelst and B.~Moons, ``Embedded deep neural network processing: Algorithmic and processor techniques bring deep learning to iot and edge devices,'' \emph{IEEE Solid-State Circuits Magazine}, vol.~9, no.~4, pp. 55--65, 2017.

\bibitem{hu2018reinforcement}
J.~Hu, H.~Zhang, and L.~Song, ``{Reinforcement learning for decentralized trajectory design in cellular UAV networks with sense-and-send protocol},'' \emph{IEEE Internet of Things Journal}, vol.~6, no.~4, pp. 6177--6189, 2018.

\bibitem{li2019robust}
S.~Li, Y.~Wu, X.~Cui, H.~Dong, F.~Fang, and S.~Russell, ``Robust multi-agent reinforcement learning via minimax deep deterministic policy gradient,'' in \emph{Proceedings of the AAAI conference on artificial intelligence}, vol.~33, no.~01, 2019, pp. 4213--4220.

\bibitem{rashid2020monotonic}
T.~Rashid, M.~Samvelyan, C.~S. De~Witt, G.~Farquhar, J.~Foerster, and S.~Whiteson, ``Monotonic value function factorisation for deep multi-agent reinforcement learning,'' \emph{Journal of Machine Learning Research}, vol.~21, no. 178, pp. 1--51, 2020.

\bibitem{son2019qtran}
K.~Son, D.~Kim, W.~J. Kang, D.~E. Hostallero, and Y.~Yi, ``Qtran: Learning to factorize with transformation for cooperative multi-agent reinforcement learning,'' in \emph{International conference on machine learning}.\hskip 1em plus 0.5em minus 0.4em\relax PMLR, 2019, pp. 5887--5896.

\bibitem{yang2018mean}
Y.~Yang, R.~Luo, M.~Li, M.~Zhou, W.~Zhang, and J.~Wang, ``Mean field multi-agent reinforcement learning,'' in \emph{International conference on machine learning}.\hskip 1em plus 0.5em minus 0.4em\relax PMLR, 2018, pp. 5571--5580.

\bibitem{shen2022multidepot}
K.~Shen, R.~Shivgan, J.~Medina, Z.~Dong, and R.~Rojas-Cessa, ``Multidepot drone path planning with collision avoidance,'' \emph{IEEE Internet of Things Journal}, vol.~9, no.~17, pp. 16\,297--16\,307, 2022.

\bibitem{yanmaz2018drone}
E.~Yanmaz, S.~Yahyanejad, B.~Rinner, H.~Hellwagner, and C.~Bettstetter, ``Drone networks: Communications, coordination, and sensing,'' \emph{Ad Hoc Networks}, vol.~68, pp. 1--15, 2018.

\end{thebibliography}

\clearpage
\appendices
\section{success rate evaluation for parameter study}
\label{app A: parameter}
\begin{figure}[h]
    \centerline{\includegraphics[scale=0.5]{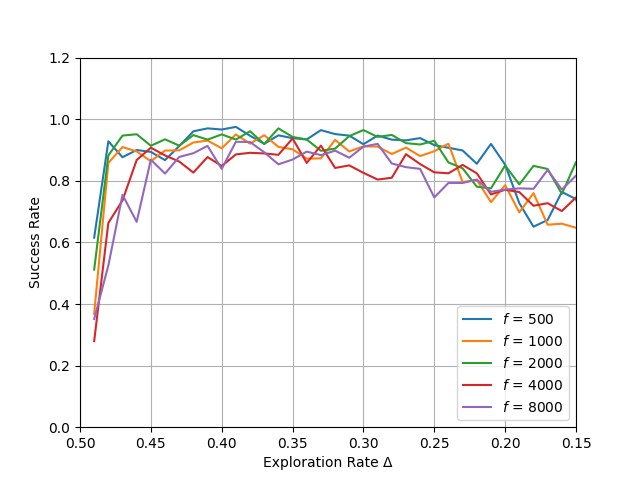}}
    \caption{The success rate under different target network update frequency \textit{f} values and exploration rates.}
    \label{fig 5:freq_success}
\end{figure}

Fig.~\ref{fig 5:freq_success} shows the success rate evaluation of different target network update frequencies \textit{f}. \textit{f}$ = 8000$ gives more stable performance similar to the execution step evaluation.

\section{success rate evaluation for task length and location study}
\label{app B: random}

\begin{figure}[h]
    \centerline{\includegraphics[scale=0.5]{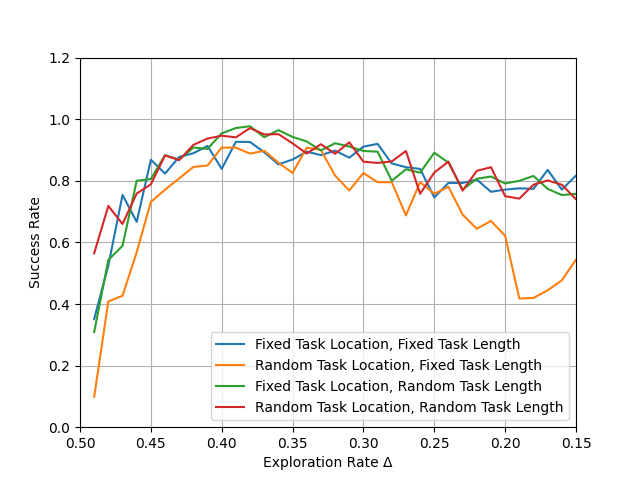}}
    \caption{The success rate under different task location and task length conditions.}
    \label{fig 6:random_success}
\end{figure}

The success rate evaluations for task length and location are shown in Fig.~\ref{fig 6:random_success}. It shows a similar result to the task execution step evaluation. Except the scenario of random task locations with fixed
task lengths, our proposed model was able to achieve around 80\% success rate for all other three scenarios. This can be explained by the similar arguments as aforementioned.

\begin{figure*}[ht!]
    \centering
    \subfloat[The success rate of finishing missions with two tasks.]{{\includegraphics[width=0.3\linewidth]{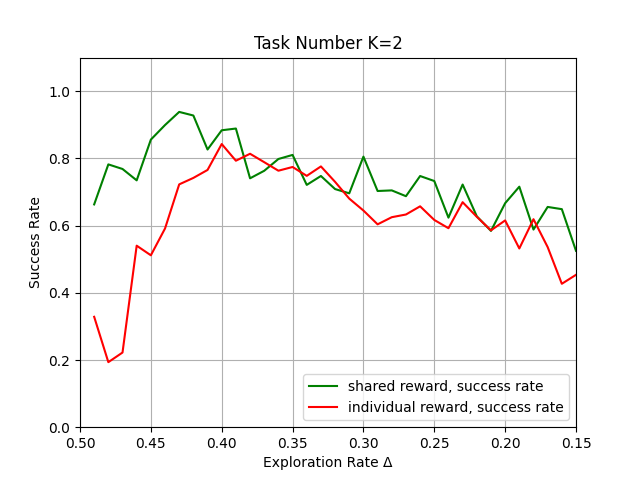} }}
    \qquad
    \subfloat[The success rate of finishing missions with four tasks.]{{\includegraphics[width=0.3\linewidth]{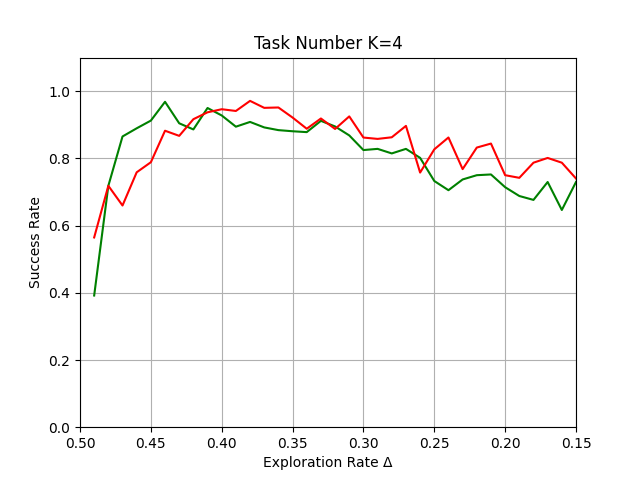} }}
    \qquad
    \subfloat[The success rate of finishing missions with six tasks.]{{\includegraphics[width=0.3\linewidth]{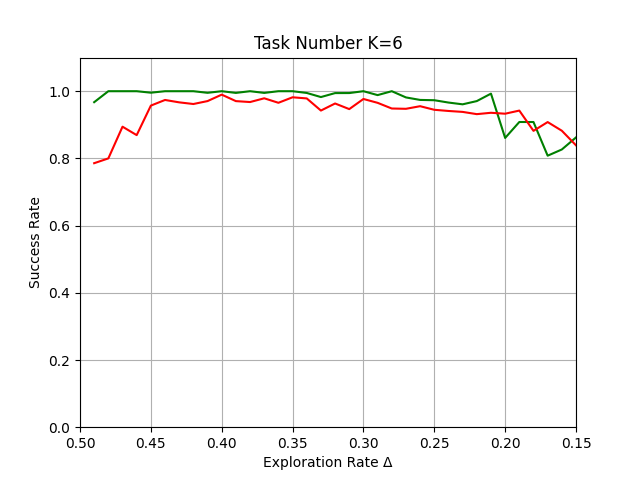} }}
    
    \medskip
    
    \subfloat[The success rate of finishing missions with eight tasks.]{{\includegraphics[width=0.3\linewidth]{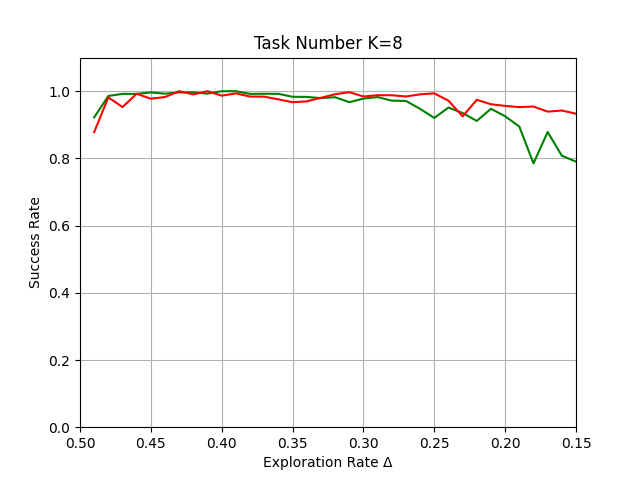} }}
    \qquad
    \subfloat[The success rate of finishing missions with ten tasks.]{{\includegraphics[width=0.3\linewidth]{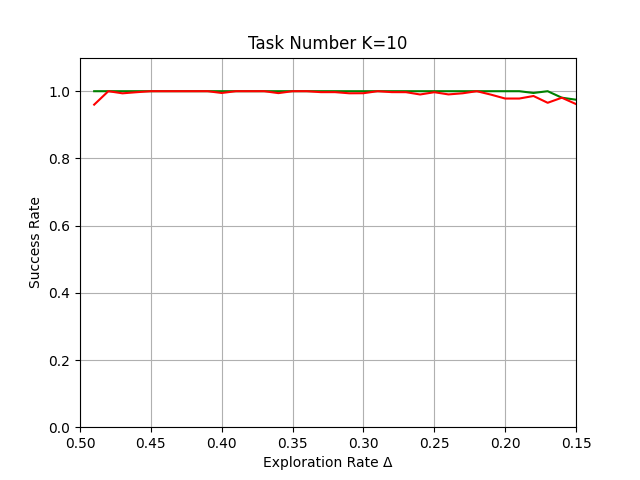} }}
    \caption{Success rate comparison of the individual reward mode and the shared reward mode on various task densities.}
    \label{fig 9:comparison_succcess}
\end{figure*}

\begin{figure*}[h!]
    \centering
    \subfloat[The average number of steps required to finish a mission with two tasks.]{{\includegraphics[width=0.3\linewidth]{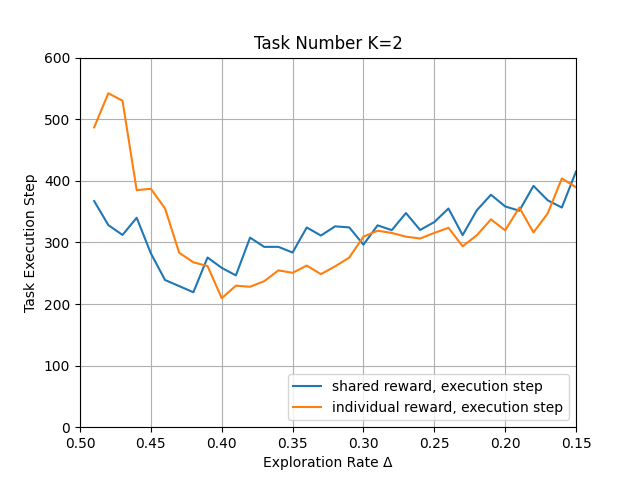} }}
    \qquad
    \subfloat[The average number of steps required to finish a mission with four tasks.]{{\includegraphics[width=0.3\linewidth]{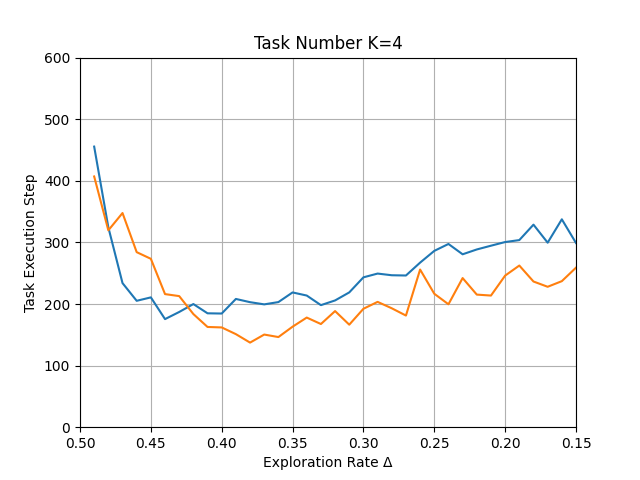} }}
    \qquad
    \subfloat[The average number of steps required to finish a mission with six tasks.]{{\includegraphics[width=0.3\linewidth]{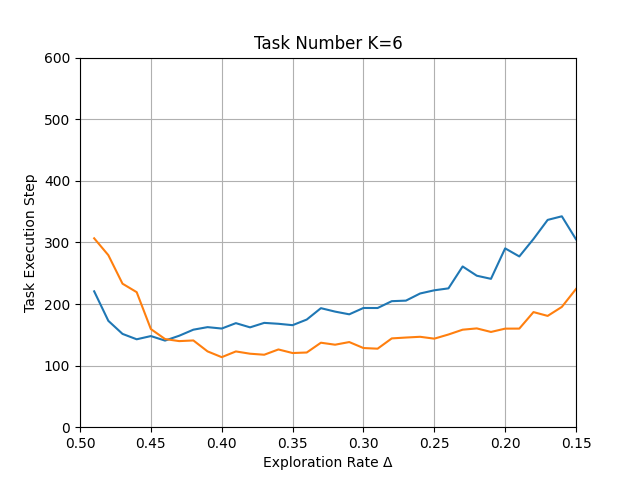} }}
    
    \medskip
    
    \subfloat[The average number of steps required to finish a mission with eight tasks.]{{\includegraphics[width=0.3\linewidth]{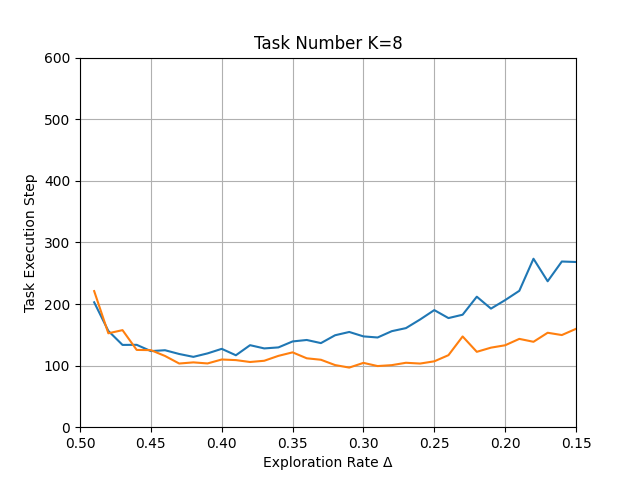} }}
    \qquad
    \subfloat[The average number of steps required to finish a mission with ten tasks.]{{\includegraphics[width=0.3\linewidth]{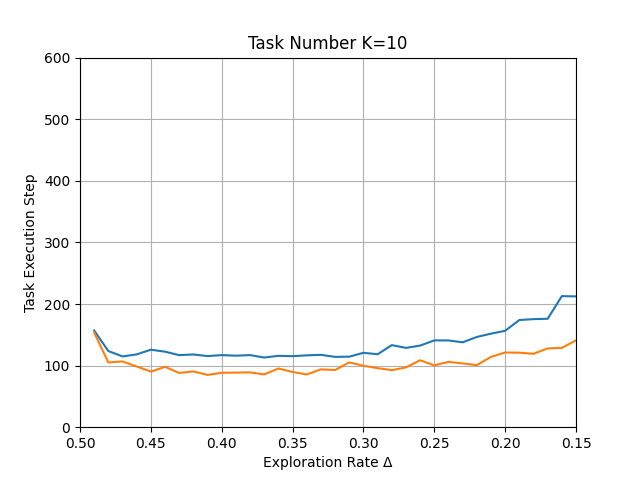} }}
    \caption{Comparison of the number of execution steps of the individual reward mode and the shared reward mode on various task densities.}
    \label{fig 9:comparison_step}
\end{figure*}

\section{Comparison of training between shared reward mode and individual reward mode}
\label{app E: comp}

\begin{figure*}[h!]
    \centering
    \subfloat[The success rate of finishing a mission in grid of size $5\times5$.]{{\includegraphics[width=0.45\linewidth]{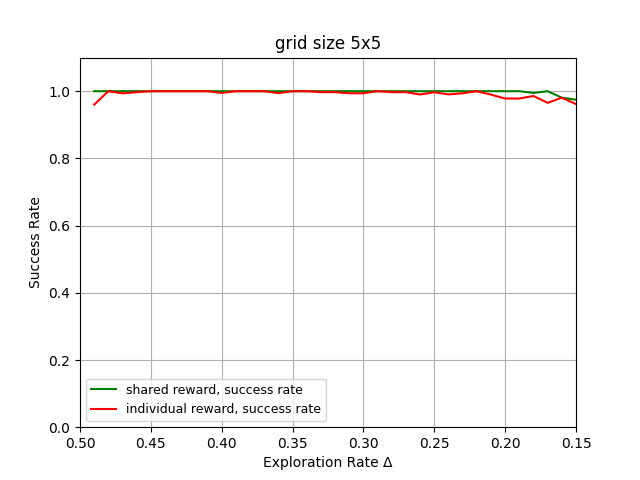} }}
    \qquad
    \subfloat[The success rate of finishing a mission in grid of size $6\times6$.]{{\includegraphics[width=0.45\linewidth]{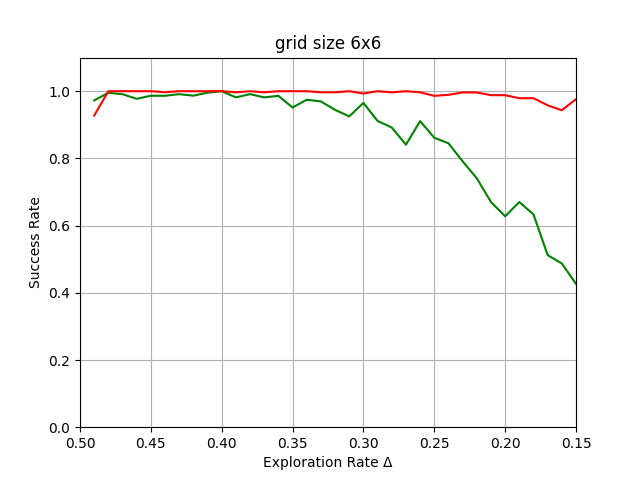} }}
    \qquad
    \subfloat[The success rate of finishing a mission in grid of size $7\times7$.]{{\includegraphics[width=0.45\linewidth]{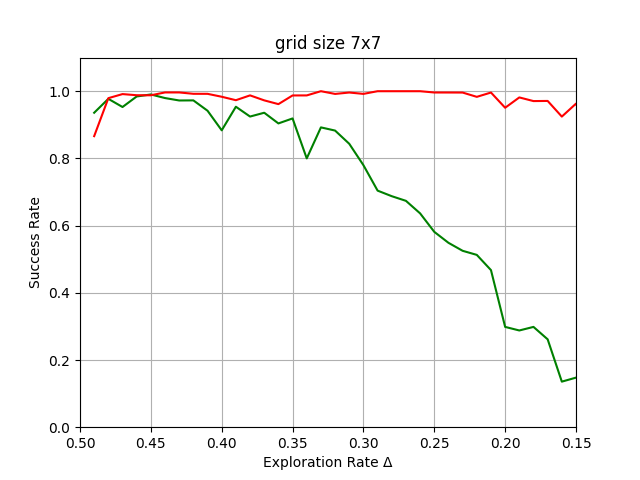} }}
    \qquad
    \subfloat[The success rate of finishing a mission in grid of size $8\times8$.]{{\includegraphics[width=0.45\linewidth]{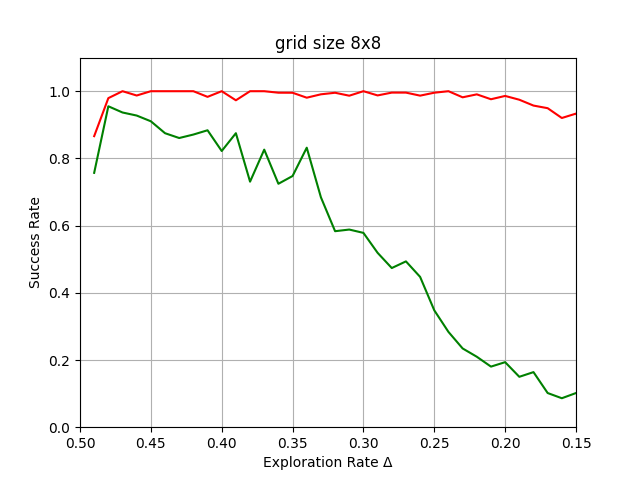} }}
    \caption{Success rate comparison of the individual reward mode and the shared reward mode under different grid sizes.}
    \label{fig 10:comparison2_success}
\end{figure*}

\begin{figure*}[h!]
    \centering
    \subfloat[The average number of steps required to finish a mission in grid of size $5\times5$.]{{\includegraphics[width=0.45\linewidth]{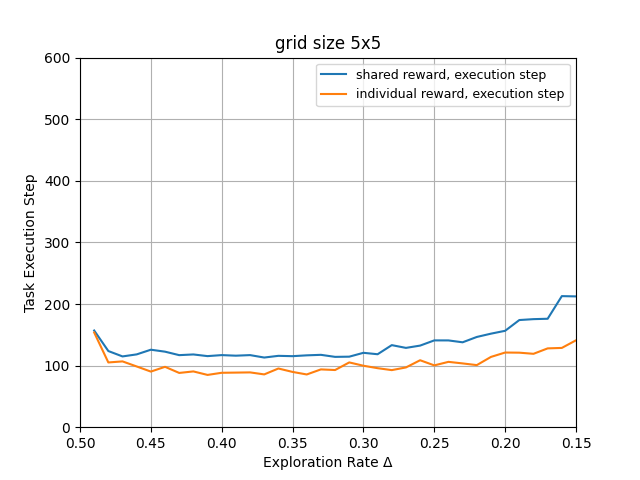} }}
    \qquad
    \subfloat[The average number of steps required to finish a mission in grid of size $6\times6$.]{{\includegraphics[width=0.45\linewidth]{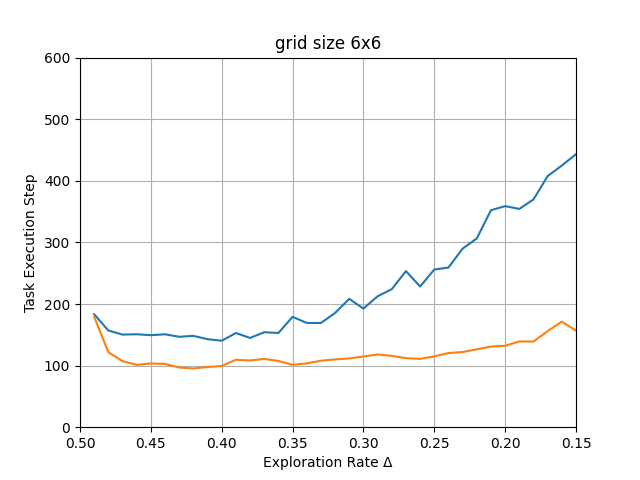} }}
    \qquad
    \subfloat[The average number of steps required to finish a mission in grid of size $7\times7$.]{{\includegraphics[width=0.45\linewidth]{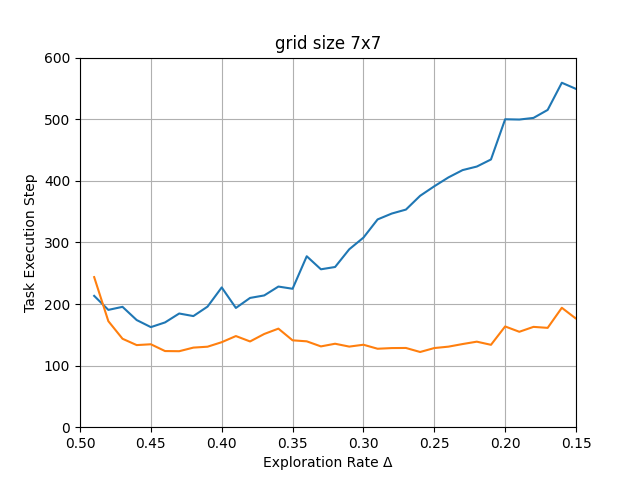} }}
    \qquad
    \subfloat[The average number of steps required to finish a mission in grid of size $8\times8$.]{{\includegraphics[width=0.45\linewidth]{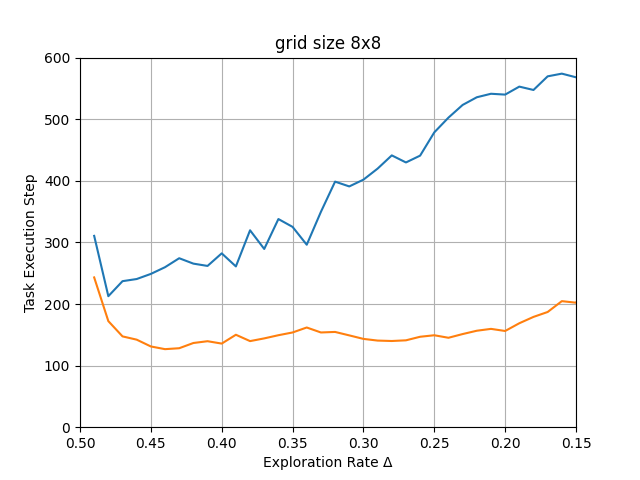} }}
    \caption{Comparison of the number of execution steps of the individual reward mode and the shared reward mode under different grid sizes.}
    \label{fig 10:comparison2_step}
\end{figure*}

Fig.~\ref{fig 9:comparison_succcess} and Fig.~\ref{fig 9:comparison_step} show the comparison of training processes of shared reward mode and individual reward mode. Fig.~\ref{fig 9:comparison_succcess} indicates that both shared reward mode and individual reward mode are able to learn the optimal policy to achieve high success rate with individual reward mode shows a relatively more stable performance. Fig.~\ref{fig 9:comparison_step} amplifies the advantage of individual reward as the required execution steps are consistently fewer than shared rewards which shows the individual reward converges faster and gives a better performance.

The training processes of both shared reward mode and individual reward mode under different grid sizes are shown in Fig~\ref{fig 10:comparison2_success} and Fig~\ref{fig 10:comparison2_step}. Fig~\ref{fig 10:comparison2_success} shows that individual reward mode is able to converge fast and maintain a high success rate approaching the end of training regardless of the change of the grid size. On the contrary, shared reward mode fails to converge when the size of the grid increases which indicates that it fails to learn an optimal policy. The results of evaluation based on required execution step aligns with the analysis. Fig~\ref{fig 10:comparison2_step} shows that for individual reward mode, the required execution step decreases as the agents are trained and it maintains a similar performance even when grid size increases. However, shared reward mode diverges and fails to learn when the grid size is big. The comparison of the training curve aligns with our analysis in the main session that individual reward mode is more robust and scalable.

\newpage

\vspace{11pt}

\vfill

\end{document}